\documentclass[twocolumn,longbibliography,superscriptaddress,amsmath,amssymb,aps,floatfix]{revtex4-2}
\usepackage{graphicx} % Required for inserting images
\usepackage[T1]{fontenc}
\usepackage{comment}
\usepackage{multirow}
\usepackage{dcolumn}% Align table columns on decimal point
\usepackage{bm}% bold math
\usepackage{braket}  % Dirac symbols
\usepackage{hyperref}% add hypertext capabilities
\hypersetup{hypertex=true,colorlinks=true,linkcolor=blue,anchorcolor=blue,citecolor=blue}
\usepackage{placeins} %Used to keep figures in their sections
\usepackage{physics}

\newcommand{\DU}[1]{\ensuremath{\ket{\downarrow \uparrow}}}

\newcommand{\avg}[1]{\left\langle #1 \right\rangle}
\newcommand{\Rcal}{\mathcal{R}}
\newcommand{\Ical}{\mathcal{I}}
\newcommand{\Vcal}{\mathcal{V}}
\newcommand{\nge}{n_{\mathrm{Ge}}}
\newcommand{\Ege}{E_{\mathrm{Ge}}}

\newcommand{\tgg}{\tilde{t}_{gg}}
\newcommand{\tge}{\tilde{t}_{ge}}
\newcommand{\teg}{\tilde{t}_{eg}}
\newcommand{\tee}{\tilde{t}_{ee}}
\newcommand{\tL}{\mathrm{L}}
\newcommand{\tR}{\mathrm{R}}
\newcommand{\phiL}{\phi_{\tL}}
\newcommand{\phiR}{\phi_{\tR}}
\newcommand{\phiRL}{\phi_{\tR\tL}}
\newcommand{\dL}{\Delta_{\tL}}
\newcommand{\dR}{\Delta_{\tR}}
\newcommand{\dRL}{\Delta_{\tR\tL}}
\newcommand{\BS}[1]{\boldsymbol{#1}}

\begin{document}

\title{Complete measurement of tunnel- and valley-coupling parameters in a silicon double quantum dot}

\author{Daniel J. King}
\thanks{These authors contributed equally to this work.}
\affiliation{Department of Physics, University of Wisconsin-Madison, Madison, Wisconsin, 53706, United States}

\author{Minyoung Kim}
\thanks{These authors contributed equally to this work.}
\affiliation{Department of Physics, University of Wisconsin-Madison, Madison, Wisconsin, 53706, United States}

\author{J. Reily}
\affiliation{Department of Physics, University of Wisconsin-Madison, Madison, Wisconsin, 53706, United States}

\author{Jonathan C. Marcks}
\affiliation{Q-NEXT, Argonne National Laboratory, Lemont, Illinois, 60439, United States}
\affiliation{Materials Science Division, Argonne National Laboratory, Lemont, Illinois, 60439, United States}
\affiliation{Pritzker School of Molecular Engineering, University of Chicago, Chicago, Illinois, 60637, United States}

\author{Mark Friesen}
\affiliation{Department of Physics, University of Wisconsin-Madison, Madison, Wisconsin, 53706, United States}

\author{Benjamin D. Woods}
\affiliation{Department of Physics, University of Wisconsin-Madison, Madison, Wisconsin, 53706, United States}

\author{M. A. Eriksson}
\affiliation{Department of Physics, University of Wisconsin-Madison, Madison, Wisconsin, 53706, United States}
\date{\today}

\begin{abstract}
Tunneling is essential in the initialization, measurement, and control of quantum dot qubits. In silicon, such tunneling connects not only the qubit states but also valley minima in the conduction band on opposite sides of the Brillouin zone, with large consequences for the quantum dot behavior. Here we present a full characterization of the intravalley and intervalley tunnel couplings, including their complex phases---the valley phases. These phases are shown to control measurable parameters, including the ratios of the gaps at anticrossings between quantum states of a double quantum dot.  The valley phases themselves evolve as a function of the quantum dot gate voltages and depend on the underlying atomic structure of the quantum well. Knowledge of the valley phases completes the picture and fills a key gap in our understanding of sample-wide variations of valley couplings and the physical parameters that depend on them, including spin-orbit coupling, valley-orbit mixing, and Landé $g$-factors.
\end{abstract}

\maketitle

Electron spins in silicon quantum dots have emerged as a prominent platform for quantum information processing, due to their small footprint, their potential for scale-up \cite{ Maurand2016p13575, Zwerver2022p184, Steinacker2025p81, Huckemann2025p868}, and their high-fidelity operations \cite{Yoneda2018p102, Yang2019p151, Lawrie2023p3617, Xue2022p343, Noiri2022p338, Mills2022peabn5130, Wu2025preprint}. 
While this materials system lends itself very well to quantum computing applications \cite{Vandersypen2019p38,Ladd2026preprint}, it is also well suited for transporting quantum information over long distances ($>$10~$\mu$m) via shuttling schemes \cite{Seidler2022p100, Struck2024p1325, Xue2024p2296, DeSmet2025p866, Matsumoto2026p1476}, and for opportunities in quantum sensing \cite{Chan2018p044017, Ryu2025p12067}. 
Despite these highlights, challenges remain for large-scale implementations due to device variability \cite{Zajac2016p054013, neyensProbingSingleElectrons2024},
especially that arising from the random-alloy disorder of the Si/SiGe heterostructure materials that host the quantum dots~\cite{Chen2021p044033, Wuetz2022p7730}.

Tunneling is a quintessentially quantum process used to manipulate quantum information. 
Owing to its nonclassical nature, tunneling depends sensitively on the details of the potential barrier between localized qubits \cite{Burkard2023p025003}. 
In quantum dots, this presents challenges for implementing high-fidelity two-qubit gates, especially in the presence of charge noise \cite{Hu2006p100501}. 
More fundamentally, however, tunneling in Si depends on the complex phase differences between the conduction-band valley eigenstates in localized dots, which are typically governed by the SiGe random alloy disorder \cite{MerrittPracticalVS}. 
Accurate characterization and command of the tunnel coupling under such challenging conditions is imperative for operating qubits at scale.

\begin{figure*}
\includegraphics{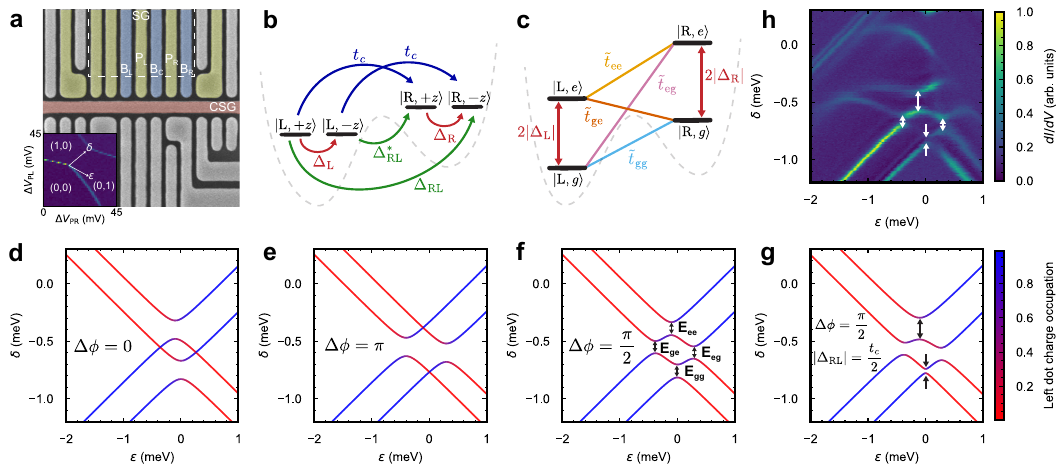}
\caption{{\bf Device and valley-phase physics.} 
{\bf a,} A false-colored, scanning electron micrograph of a device lithographically identical to the one used in the experiment.  Dots are formed using plunger gates $\mathrm{P}_\mathrm{L}$ and $\mathrm{P}_\mathrm{R}$; $\mathrm{B}_\mathrm{L}$, $\mathrm{B}_\mathrm{C}$, and $\mathrm{B}_\mathrm{R}$ label the barrier gates, and the dashed white line shows the $\mathrm{SG}$ screening gate, located beneath the visible gate layer. \emph{Inset:} charge-stability diagram, showing the $\varepsilon$ and $\delta$ tuning axes, with respect to the $(0,0)\text{-}(1,0)\text{-}(0,1)$ triple point. 
{\bf b,} Illustration of couplings in a DQD in the $\pm z$-valley basis of the four-level model, where $t_{c}$ is the valley-conserving tunnel coupling, $\Delta_\mathrm{L}$ and $\Delta_\mathrm{R}$ are intradot-intervalley couplings, and $\Delta_\mathrm{RL}$ is the interdot-intervalley coupling. 
{\bf c}, Illustration of couplings in the ground/excited-valley basis after diagonalization of $\Delta_\mathrm{L}$ and $\Delta_\mathrm{R}$, where the ground and excited valleys are separated by the valley energy splittings $E_{v,i} = 2|\Delta_{i}|$.
Note that both intravalley $(\tilde{t}_{gg}, \tilde{t}_{ee})$ and intervalley $(\tilde{t}_{eg}, \tilde{t}_{ge})$ tunnel couplings are present, and both play an important role in this work.
{\bf d-g,} Example spectra of the four-level model, with valley phase differences $\Delta\phi$ of ({\bf d}) 0, ({\bf e}) $\pi$,  and ({\bf f, g}) $\frac{\pi}{2}$. 
In ({\bf d-f}), $\Delta_\mathrm{RL} = 0$, while in ({\bf g}), $\Delta_\mathrm{RL} \neq 0$, leading to $E_{ee}\gg E_{gg}$. 
{\bf h,} A DAXS spectrum yielding a direct measurement of the single-electron energy dispersion, whose low-energy states closely resemble the four-level model spectrum in ({\bf g}).}
\label{FIG1}
\end{figure*}

In this paper, we perform experimentally a full parametric characterization of the intravalley and intervalley tunnel couplings in a Si double quantum dot, including their complex phases (known as \emph{valley phases} \cite{Friesen2007p115318}). 
Theoretically, we clarify the relation between the various valley phases emerging in this system, particularly the phase corresponding to the interdot-intervalley coupling, whose importance has been previously overlooked.
Our experimental approach leverages a recently developed technique, known as delta-axis spectroscopy (DAXS), which provides complete mappings of the double-dot energy dispersion via baseband methods \cite{reilydaxs}. 
By enhancing the resolution of DAXS measurements at the energy-level anticrossings, we obtain energy spectra with sufficient detail to permit the simultaneous fitting of six different Hamiltonian parameters.  These include the valley coupling parameters, which are complex, and their phases are shown to depend sensitively on the local atomic distribution of Ge in the quantum well.
Together, these results provide a clear picture of how tunnel couplings and energy splittings emerge from the underlying microscopic physics of a Si double quantum dot.

\section*{Double quantum dot experiment}

We measure tunnel couplings, valley splittings, and valley phases in an Intel Tunnel Falls Si/SiGe quantum dot device, lithographically identical to the one shown in the scanning electron micrograph in Fig.~\ref{FIG1}a \cite{neyensProbingSingleElectrons2024, george12SpinQubitArraysFabricated2025}. 
Here, 1.7\% Ge is incorporated into the Si quantum well to enhance the average valley splitting \cite{MerrittPracticalVS,Marcks2025}. A double quantum dot (DQD) is formed using the $\mathrm{P}_\mathrm{L}$ and $\mathrm{P}_\mathrm{R}$ plunger gates. Tunnel couplings between the dots and to adjacent reservoirs are controlled by the $\mathrm{B}_\mathrm{L}$, $\mathrm{B}_\mathrm{C}$, and $\mathrm{B}_\mathrm{R}$ barrier gates, along with the $\mathrm{SG}$ screening gate located beneath the plunger and barrier gates. The DQD is tuned near the $(0,0)\text{-} (1,0) \text{-}(0,1)$ charge-occupation triple point, shown in the charge stability diagram in the inset of Fig.~\ref{FIG1}a, to probe the single-electron physics.

As we discuss below, the critical new experimental step to extract all the valley coupling magnitudes and phases is the simultaneous measurement of all four anticrossing gaps between the ground and excited valley states of the two dots. Knowledge of both the gap locations in detuning energy and the sizes of the gaps themselves is important and corresponds to six independent pieces of information. Previously, three of these anticrossings have been characterized, which is not sufficient to extract valley phases. The ground state anticrossing is the polarization line, and its width has been measured in dc charge sensing~\cite{DiCarlo2004,Simmons_2009} and the gap itself has been observed in DAXS~\cite{yoo_daxs_at_01_10_00}. Microwave spectroscopy has been used to measure the ground state anticrossing and the next two anticrossings, between the first excited valley state of one dot and the ground valley state of the other dot, and \emph{vice versa}~\cite{borjansProbingVariationIntervalley2021a}.  Here we use DAXS to measure all four anticrossings in a two step procedure described below: first, DAXS is used to map out the energy vs.\ detuning diagram to locate all four anticrossings.  Second, high-resolution DAXS measurements are acquired to enable fitting and determination of the intradot and interdot valley couplings.

\section*{Tunnel couplings and valley couplings}

The low-energy single-electron physics of the DQD can be understood from the four-level model illustrated in Fig.~\ref{FIG1}b. Here, a single spatial orbital ($\mathrm{L}/\mathrm{R}$) is considered in each dot. In addition, each spatial orbital has a valley degree of freedom ($\pm z$) due to the two-fold degenerate valley minima in the conduction band of the biaxially strained Si quantum well \cite{Zwanenburg2013}. The various level couplings are illustrated by arrows in Fig.~\ref{FIG1}b. These include valley-conserving tunnel coupling $t_c$ as well as intradot and interdot valley-coupling matrix elements defined as $\Delta_{ij} = \mel{i,-z}{H}{j,+z}$, where $i,j \in \left\{\mathrm{R},\mathrm{L}\right\}$. Importantly, the valley-couplings are complex, $\Delta_{ij} = |\Delta_{ij}| e^{i \phi_{ij}}$, where $\phi_{ij}$ is the corresponding valley phase. We denote $\Delta_{ii}$ and $\phi_{ii}$ by $\Delta_{i}$ and $\phi_{i}$, respectively, where $2|\Delta_i|$ is the routinely measured valley splitting of a single dot \cite{Marcks2025, Chen2021p044033, dodson_how_2022}.
Here, we emphasize the inclusion of $\Delta_\mathrm{RL}$, which we call the interdot-intervalley coupling. 
The various valley couplings are all randomized by the SiGe alloy disorder \cite{MerrittPracticalVS}, and their statistics are determined by the Ge concentration profile of the heterostructure and the low-energy wavefunctions, as discussed in the Supplementary Materials \cite{SM}.

The dominant level couplings are typically the intradot-intervalley couplings, $\Delta_\mathrm{L}$ and $\Delta_\mathrm{R}$. Therefore, it is useful to work in a ground/excited-valley basis that diagonalizes this coupling, leading to the transformed level diagram shown in Fig.~\ref{FIG1}c.  This basis is both convenient for theory and also provides the typical experimental vocabulary: the ground ($g$) and excited ($e$) states of each dot are split by their respective valley splittings, $E_{v,i} = 2|\Delta_{i}|$, and these are the commonly named \emph{valley states} of each quantum dot.

As shown in Fig.~\ref{FIG1}c, in this basis there exist both intravalley $(\tilde{t}_{gg}, \tilde{t}_{ee})$ and intervalley $(\tilde{t}_{eg}, \tilde{t}_{ge})$ tunnel couplings between the dots.  Importantly, these couplings can be written in terms of the valley phases as follows 
(see \cite{SM} for a derivation):
\begin{align}
    \tilde{t}_{gg} &= t_{c} \cos\left(\Delta\phi/2\right) - |\Delta_\mathrm{RL}| \cos\left(\Delta\phi/2 - \Delta\phi_\mathrm{RL}\right), \label{tgg} \\
    \tilde{t}_{ee} &= t_{c} \cos\left(\Delta\phi/2\right) + |\Delta_\mathrm{RL}| \cos\left(\Delta\phi/2 - \Delta\phi_\mathrm{RL}\right), \label{tee} \\
    \tilde{t}_{ge} &= t_{c} \sin\left(\Delta\phi/2\right) - |\Delta_\mathrm{RL}| \sin\left(\Delta\phi/2 - \Delta\phi_\mathrm{RL}\right), \label{tge} \\
    -\tilde{t}_{eg} &= t_{c} \sin\left(\Delta\phi/2\right) + |\Delta_\mathrm{RL}| \sin\left(\Delta\phi/2 - \Delta\phi_\mathrm{RL}\right), \label{teg}
\end{align}
where $\Delta \phi = \phi_\mathrm{L} - \phi_\mathrm{R}$ is the valley phase difference and $\Delta \phi_\mathrm{RL} = \phi_\mathrm{RL} - \phi_\mathrm{R}$. Here, without loss of generality, we choose a convention where $0 \leq \Delta\phi \leq \pi$ and $-\pi < \Delta\phi_\mathrm{RL} \leq \pi$. 
Eqs.~\eqref{tgg}-\eqref{teg} show that the relative strengths of the intravalley and intervalley tunnel couplings are determined by the valley phases, $\Delta\phi$ and $\Delta\phi_\mathrm{RL}$, and the ratio $|\Delta_\mathrm{RL}|/t_{c}$. 

The valley-phase modulations of the tunnel couplings in Eqs. (\ref{tgg})-(\ref{teg}) are  easily visualized in the detuning spectrum of the DQD, as illustrated for several different system parameters in Fig.~\ref{FIG1}d-g. 
In these detuning diagrams, the $\mathrm{L}$ and $\mathrm{R}$ dot energy levels move up and down, respectively, with increasing detuning $\varepsilon$, yielding four anticrossings.  Their corresponding gaps $E_{\mu \nu}$ are labeled in Fig.~\ref{FIG1}f, where $\mu,\nu \in \{g,e\}$. 
Each gap is primarily determined by its corresponding tunnel coupling, $E_{\mu \nu} \approx 2|\tilde{t}_{\mu \nu}|$. 

The importance of each term is easily visualized by first considering Fig.~\ref{FIG1}d-f, where the interdot-intervalley coupling is set to zero ($\Delta_\mathrm{RL} = 0$), and the valley phase difference $\Delta \phi$ is set respectively to $0$, $\pi$, and $\pi/2$. Corresponding to Eqs.~(\ref{tgg})-(\ref{teg}), we then see that the relative tunnel coupling strengths are controlled solely by the valley phase difference $\Delta\phi$, leading to $\tilde{t}_{gg}\! =\! \tilde{t}_{ee}$, and $\tilde{t}_{ge}\! =\! -\tilde{t}_{eg}$. 
For the case $\Delta\phi =0$ (Fig.~\ref{FIG1}d), the intervalley tunnel couplings are extinguished ($\tilde{t}_{eg}\!=\! \tilde{t}_{ge}\! =\! 0$), leading to $E_{eg} \! = \! E_{ge}\! =\! 0$. 
In the opposite limit of $\Delta\phi = \pi$ (Fig.~\ref{FIG1}e), we find $E_{gg} \!=\! E_{ee} \!=\! 0$. 
For the intermediate phase difference of $\Delta\phi = \pi/2$ (Fig.~\ref{FIG1}f), all the gaps are equal. 
Importantly, we obtain $E_{gg} = E_{ee}$ and $E_{ge} = E_{eg}$, for all values of $\Delta\phi$, when $\Delta_\mathrm{RL} = 0$.

This symmetric behavior no longer appears once the interdot-intervalley coupling $\Delta_\mathrm{RL}$ is introduced, as is evident from Eqs.~(\ref{tgg})-(\ref{teg}). This situation is illustrated in Fig.~\ref{FIG1}g, where $|\Delta_\mathrm{RL}| = t_{c}/2$ and $E_{ee} \gg E_{gg}$.  We stress that $\Delta_\mathrm{RL}$ is typically ignored in models of Si DQDs~\cite{culcerInterfaceRoughnessValleyorbit2010,saraivaIntervalleyCouplingInterfacebound2011, borjansProbingVariationIntervalley2021a,Buterakos2021,woodsNegativeExchangeInteraction2025}. We find below that $\Delta_\mathrm{RL}$ is an essential ingredient in fitting our experimental data.

\begin{figure*}
\includegraphics{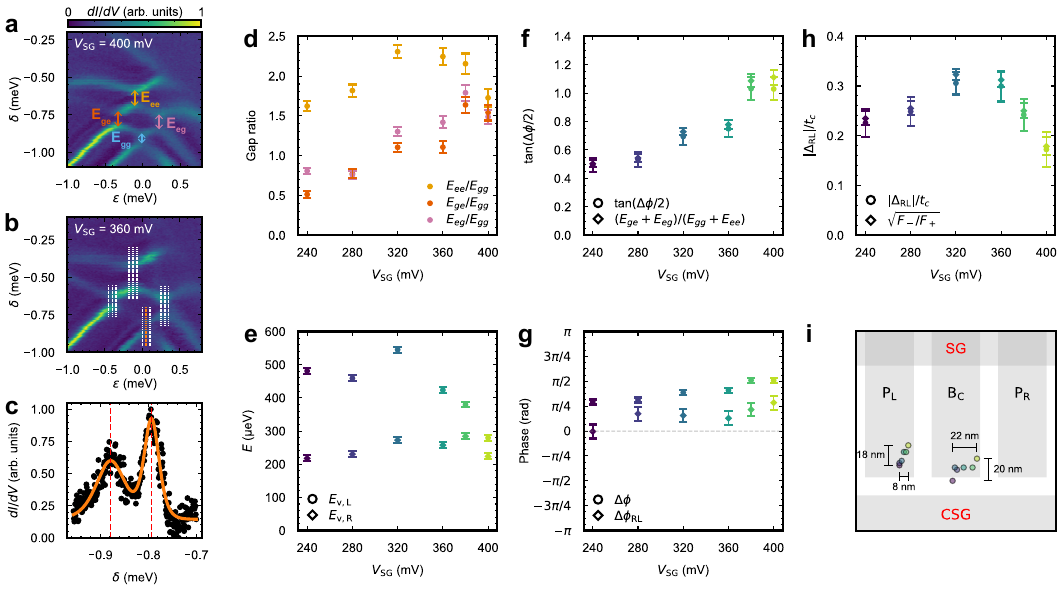}
\caption{
{\bf Extraction of Hamiltonian parameters and their evolution with the SG gate voltage, which shifts the positions of the quantum dots.}
{\bf a-b,} DAXS spectra for two SG gate voltages. The four anticrossing energy gaps are labeled in ({\bf a}). 
The dotted lines in (\textbf{b}) indicate the $\varepsilon$ values and the ranges in $\delta$ over which high-resolution one-dimensional DAXS measurements are acquired.
{\bf c,} An example high-resolution DAXS scan, corresponding to the orange line segment in ({\bf b}). 
The best-fit curve is overlaid in orange, with the extracted peak locations in $\delta$ denoted by vertical dashed lines.
{\bf d,} Anticrossing energy gap ratios $E_{\mu\nu}/E_{gg}$. 
{\bf e,} Extracted valley splittings $E_{v,i}$ of the two dots. 
$E_{v,L}$ varies by a factor of $\sim$2, due to the shifting dot position. 
{\bf f,} Comparison of the quantities $\tan(\Delta \phi/2)$ and $(E_{ge} +E_{eg})/(E_{gg} + E_{ee})$, where $\Delta\phi$ is obtained by fitting to the four-level model. The remarkable agreement between these two quantities demonstrates that $\Delta\phi$ is encoded in the four gaps.
{\bf g,} Extracted valley phase differences, $\Delta \phi$ and $\Delta \phi_\mathrm{RL}$.
Significant changes to both phases are observed, which we attribute to the shifting dot positions. 
{\bf h,} Comparison of the tunnel coupling ratio $|\Delta_\mathrm{RL}|/t_{c}$ and the dimensionless gap ratio $\sqrt{F_-/F_+}$, which is defined in the main text. 
As in ({\bf f}), the good agreement between these quantities demonstrates that important Hamiltonian parameters are encoded in the gap ratios. 
Note that $|\Delta_\mathrm{RL}|/t_{c}$ is non-negligible over the entire $V_\text{SG}$ range, demonstrating that interdot-intervalley coupling is important for the quantum dot physics and that this coupling information also is encoded in the anticrossing gaps. 
{\bf i,} Estimated dot positions, based on capacitance measurements and electrostatic calculations. The data in ({\bf e}-{\bf h}) are color-coded to correspond with the estimated dot positions in ({\bf i}).
}
\label{FIG2}
\end{figure*}

\section*{Measuring the energy dispersion}

In Fig.~\ref{FIG1}h, we probe using delta-axis spectroscopy (DAXS)~\cite{reilydaxs} the energy dispersion and valley physics of the DQD formed in Fig.~\ref{FIG1}a. DAXS is a recently developed technique that directly maps the energy states of a DQD as a function of detuning, yielding energy dispersions, exemplified in Fig.~\ref{FIG1}h, that closely resemble the theoretical dispersions shown in Fig.~\ref{FIG1}d-g. 
We perform DAXS here by applying periodic, square wave voltage pulses along the $\delta$ tuning axis, identified in the inset of Fig.~\ref{FIG1}a. In contrast to the detuning axis $\varepsilon$, which moves the quantum dot energies in opposite directions, voltage pulses along $\delta$ shift the energies of both dots together.
Similar to detuning pulses, pulsing $\delta$ requires simultaneous voltage changes to be applied to both plunger gates. 
Each pulse can induce an electron to load or unload when the chemical potential of a DQD state crosses the Fermi level of a neighboring reservoir. 
This change in DQD charge occupation is detected by measuring the time-averaged current through the charge sensor, shown in the lower half of the device in Fig.~\ref{FIG1}a. 
Importantly, these $\delta$ pulses shift the energies of the DQD states globally with respect to the Fermi level, allowing a direct measurement of the energy spectrum, for arbitrary detuning values.

\begin{figure*}
\includegraphics{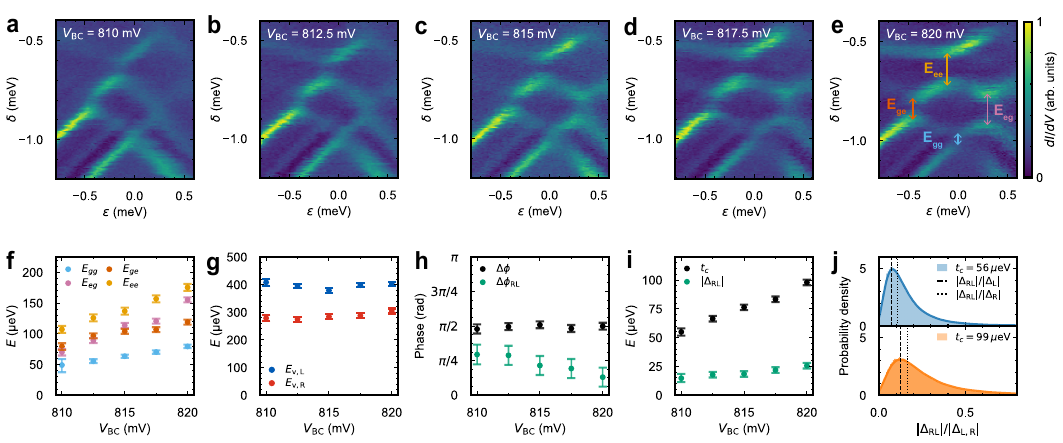}
\caption{{\bf The evolution of interdot couplings as a function of barrier-gate voltage.}  
{\bf a-e,} DAXS spectra showing the widening of the anticrossing energy gaps as a function of barrier-gate voltage, $V_\mathrm{BC}$, ranging from ({\bf a}) 810~mV to ({\bf e}) 820~mV, in steps of 2.5~mV. The energy gaps of the four anticrossings are labeled in ({\bf e}). 
{\bf f,} Energy gap values for all five DAXS measurements. 
{\bf g,} Extracted valley splittings $E_{\mathrm{v,L}}$ and $E_{\mathrm{v,R}}$ for the left and right dots, respectively. 
{\bf h,} Extracted valley phase difference $\Delta\phi$ and interdot-intervalley phase difference $\Delta\phi_\mathrm{RL}$. Note that $\Delta\phi_\mathrm{RL}$ changes significantly even for the small changes in barrier gate voltage studied here.
{\bf i,} Extracted valley-conserving tunnel coupling $t_c$ and interdot-intervalley coupling $|\Delta_\mathrm{RL}|$. 
{\bf j,} Probability distribution of the ratio $|\Delta_\mathrm{RL}|/|\Delta_\mathrm{i}|$, where $i\in\{L,R\}$. 
The two distributions correspond to $t_c =56$ and $99~\mu\text{eV}$, the experiment tunnel couplings corresponding to the first and last $V_\text{BC}$ values, as shown in ({\bf i}). Vertical dotted and dashed lines mark the measured ratios $|\Delta_\mathrm{RL}|/|\Delta_\mathrm{L}|$ and $|\Delta_\mathrm{RL}|/|\Delta_\mathrm{R}|$ respectively, showing that the values of these quantities extracted from experiment are not far from the mode of the theoretically calculated probability distribution.}
\label{FIG3}
\end{figure*}

The clear correspondence between the DAXS dispersion in Fig.~\ref{FIG1}h and the theoretical dispersions of Figs.~\ref{FIG1}d-g enables a detailed analysis of the DQD Hamiltonian parameters, illustrated in Fig.~\ref{FIG1}b and c, including the tunnel coupling $t_{c}$, the magnitudes of the intervalley couplings $(|\Delta_\mathrm{L}|, |\Delta_\mathrm{R}|, |\Delta_\mathrm{RL}|)$, and the relative valley phases $(\Delta\phi, \Delta\phi_\mathrm{RL})$. 
We perform this parameter extraction as follows (see \cite{SM} for additional details).
First, we acquire a global DAXS spectrum, like the one shown in Fig.~\ref{FIG1}h, from which we determine the approximate locations and sizes of the four anticrossings. 
Second, we perform high-resolution, one-dimensional DAXS scans for three different $\varepsilon$ values near each anticrossing, to more accurately determine the energy levels. 
Each high-resolution scan is repeated ten times, both to improve the accuracy of the extracted energy levels, and to help establish the energy uncertainty of each level. 
Third, we fit the tunneling parameters $\tilde{t}_{gg}, \tilde{t}_{ee}, \tilde{t}_{eg}$, and $\tilde{t}_{ge}$, and the valley splittings $2|\Delta_\mathrm{L}|$ and $2|\Delta_\mathrm{R}|$ of the four-level Hamiltonian discussed in Fig.~\ref{FIG1}c, using the energy levels extracted from the high-resolution scans. 
Finally, $\tilde{t}_{gg}, \tilde{t}_{ee},\tilde{t}_{eg}$, and $\tilde{t}_{ge}$ are inverted to give $t_{c}, |\Delta_\mathrm{RL}|, \Delta\phi$, and $\Delta\phi_\mathrm{RL}$ from Eqs.~(\ref{tgg})-(\ref{teg}).  
As discussed in \cite{SM}, the estimated uncertainties of the extracted energy levels from the high-resolution scans are determined for the fitted Hamiltonian parameters using a Monte Carlo error propagation technique. 
In \cite{SM}, we also quantify uncertainties in the Hamiltonian parameters arising from interactions with higher energy levels, which are not included in the four-level model, and alterations of the confinement potential shape with changing $\delta$.
(Some of these high-energy bands can be seen in Fig.~\ref{FIG1}h.) Importantly, these uncertainties are included in the error bars for the results presented in Figs.~\ref{FIG2} and \ref{FIG3}, as described in~\cite{SM}.

\section*{System evolution with screening-gate voltage and dot position}

We now perform DAXS measurements while systematically varying different gate voltages, to study their effect on the Hamiltonian parameters. 
We first vary the screening gate voltage $V_{\text{SG}}$, in order to shift the quantum dot locations relative to the atomic distributions of Ge and Si in the quantum well~\cite{MerrittPracticalVS, Marcks2025}.  For each $V_{\text{SG}}$, the finger gate voltages are retuned in order to remain at the correct triple point and to maintain the correct tunnel rates from the dots to the reservoirs, which is important for state visibility in DAXS.
Figs.~\ref{FIG2}a, b show two example DAXS spectra for $V_{\text{SG}} = 400$~mV and $360$~mV.
In Fig.~\ref{FIG2}a the anticrossing gaps are labeled, while in Fig.~\ref{FIG2}b, the locations of the high-resolution DAXS measurements are indicated. 
Fig.~\ref{FIG2}c shows a typical high-resolution measurement, where the peaks are fit to derivatives of the Fermi-Dirac distribution. Following the procedure described above, for each of six values of $V_{\text{SG}}$, we fit these peak locations in order to extract the Hamiltonian parameters, yielding the results shown in Figs.~\ref{FIG2}d-h. 
The corresponding dot positions are estimated using the electrostatic simulation procedure described in Methods, yielding the results shown in Fig.~\ref{FIG2}i, where the dots are found to move tens of nanometers over the full $V_\mathrm{SG}$ range.  Note that the positive voltage on the barrier gate $\text{B}_\text{C}$ appears to have pulled the left dot to the edge of $\text{P}_\text{L}$ and the right dot past the edge of $\text{P}_\text{R}$ and in fact under $\text{B}_\text{C}$ itself.  Nonetheless, we can still use the plungers for the DAXS pulses, and, as shown in the next section below, the voltage on $\text{B}_\text{C}$ remains effective as expected at controlling the interdot tunnel rates.

Varying $V_\text{SG}$ induces significant changes that are visible both directly in the DAXS spectra and in the extracted Hamiltonian parameters.  Fig.~\ref{FIG2}d reports the ratio of the three anticrossing gaps involving an excited state to the ground state anticrossing gap, $E_{\mu\nu}/E_{gg}$, all of which vary significantly as the dots move.
Fig.~\ref{FIG2}e reports the evolution of the valley splittings $E_{v,i}$ over this same range in $V_\text{SG}$, showing a nearly two-fold decrease in the left-dot valley splitting, and much less change in the right-dot valley splitting. 
All of these changes in gaps and valley splittings are consistent with expectations from previous work for cases in which the quantum dots move relative to the atomic alloy in the quantum well \cite{Hollmann2020p034068,Chen2021p044033,Wuetz2022p7730,limaValleySplittingDepending2024, Volmer2024,Marcks2025,Volmer2026}. 

The anticrossing gaps in the energy dispersion, $E_{\mu\nu}$, are physically intuitive quantities, and we now show that the gap ratios in Fig.~\ref{FIG2}d encode information about the valley couplings and valley phases in a simple way.  We emphasize, however, that knowledge of all four gaps are needed to actually extract the couplings and phases.
The encoding is best understood by deriving appropriate dimensionless quantities in the low-tunneling limit (or equivalently, the large-valley-splitting limit), where the anticrossings are well-separated and $E_{\mu \nu} \approx 2|t_{\mu \nu}|$. 
In this limit, Eqs.~(\ref{tgg})-(\ref{teg}) can be rearranged to give $\tan(\Delta \phi/2) \approx (E_{ge} +E_{eg})/(E_{gg} + E_{ee})$, which describes the ratio between the intermediate and upper/lower anticrossings. 
In Fig.~\ref{FIG2}f, we plot both of these quantities, $\tan(\Delta \phi/2)$ and $(E_{ge} +E_{eg})/(E_{gg} + E_{ee})$.  Here, $\Delta \phi$ is extracted from a rigorous fit to the four-level model, and $E_{\mu\nu}$ are the gap ratios shown in Fig.~\ref{FIG2}d, multiplied by $E_{gg}$.  
The differences between the two quantities in Fig.~\ref{FIG2}f are smaller than the error bars, and can be attributed to interactions between the anticrossings, away from the limit of large valley splittings or small tunnel couplings.
Note that we have adopted a sign convention in the choice of $E_{\mu \nu} =\pm 2t_{\mu \nu}$, as discussed in \cite{SM}.

Figure~\ref{FIG2}g shows extracted values of $\Delta \phi$, which increase from approximately $5\pi/16$ to $\pi/2$ within the SG voltage range. Similar to the behavior observed in Fig.~\ref{FIG2}d, the changes in $\Delta\phi$ are caused by the spatially varying alloy disorder sampled by the dots as their positions shift.
We also plot fitting results for $\Delta\phi_\mathrm{RL}$ in Fig.~\ref{FIG2}g, showing significant variations with $V_{\text{SG}}$. 

The phases and couplings plotted in Fig.~\ref{FIG2} determine the physical behavior of the DQD.  As an example, the DAXS plot shown in Fig.~\ref{FIG2}a corresponds to $V_{\text{SG}} = 400~\text{mV}$.  Our valley phase and valley coupling extraction for this $V_{\text{SG}}$ yields results consistent with the special case of $(\Delta \phi,\Delta\phi_\mathrm{RL})= (\pi/2,\pi/4)$ that was reported above in Fig.~\ref{FIG1}g.
For this case, the valley phases cause the $\Delta_\mathrm{RL}$ terms in Eqs.~\eqref{tgg} and \eqref{tee} to have maximum effect, while those terms in Eqs.~(\ref{tge}) and (\ref{teg}) have vanishing effect.  As a result, $E_{eg} = E_{ge}$, whereas $(E_{ee} - E_{gg})/|\Delta_\mathrm{RL}|$ is maximized as a function of the valley phases. Thus, the DAXS measurements enable extraction of the valley phases, and those phases in turn through Eqs.~\eqref{tgg}-\eqref{teg} enable direct checks back to the raw data. 

As a second example of gap ratios encoding Hamiltonian parameters of interest, we consider the interdot-intervalley coupling parameter $|\Delta_\mathrm{RL}|$, which can be expressed as $|\Delta_\mathrm{RL}|/t_c \approx \sqrt{F_-/F_+}$ in the large-valley-splitting (or low-tunnel coupling) limit, where we define $F_\pm = (E_{ee} \pm E_{gg})^2 + (E_{ge} \pm E_{eg})^2$.
In Fig.~\ref{FIG2}h, we plot fitting results for $|\Delta_\mathrm{RL}|/t_c$ and the dimensionless gap parameter $\sqrt{F_-/F_+}$. 
Similar to Fig.~\ref{FIG2}f, we again observe excellent agreement between the two quantities, with small differences arising from interactions between the anticrossings.
In particular, we note that $|\Delta_\mathrm{RL}|/t_c > 0.17$ is appreciable over the entire $V_{\text{SG}}$ range.
This explains why $E_{ee}/E_{gg}$ and $E_{ge}/E_{eg}$ often deviate significantly from their limiting value of $1$, which is only valid in the case of $\Delta_\mathrm{RL} = 0$. 
The results of this section underscore the importance of being able to measure all four anticrossing gaps using DAXS, as all four are needed to extract the valley phases in the general case, corresponding to the real experimental device here, in which $|\Delta_\mathrm{RL}|/t_{c}$ is non-negligible.

\section*{System evolution with barrier-gate voltage}

We now study the dependence of the Hamiltonian parameters on the $\text{B}_\text{C}$ barrier gate voltage.
DAXS spectra are reported in Figs.~\ref{FIG3}a-e for five values of $V_\text{BC}$, increasing from left to right, with $V_\text{SG}$ fixed at 380~mV. 
In Fig.~\ref{FIG3}f, we plot the resulting energy gaps of the anticrossings, which increase systematically as $V_\text{BC}$ increases, as is expected even for such small changes in barrier gate voltage, since tunnel rates are exponentially sensitive to gate voltages. 
The extracted values for the valley splittings, shown in Fig.~\ref{FIG3}g, are nearly constant over this same range.
This behavior also is expected, since the dot positions should not shift significantly with relatively small changes in $V_\mathrm{BC}$.  This nearly constant valley splitting, however, is in stark contrast with Fig.~\ref{FIG2}e, where the dot positions were not stationary.

The valley phase differences $\Delta\phi$ and $\Delta\phi_\mathrm{LR}$ shown in Fig.~\ref{FIG3}h show very different behavior from each other, and that difference arises from the different physical roles these two phases play in the DQD device physics. $\Delta\phi \approx \pi/2$ remains nearly constant over the entire range of $V_\mathrm{BC}$, while $\Delta\phi_\mathrm{RL}$ varies significantly. 
Similar to the valley splitting results, the stability of $\Delta\phi$ can be explained by the absence of dot motion: because the dot wavefunctions do not experience variations in alloy disorder, $\Delta\phi$ does not change. 
However, changes in $V_\text{BC}$ directly change the tunnel barrier height, which in turn has a large affect on the wavefunction overlap between the dots. It is this overlap region, and specifically the atomic distribution of Ge in that region, that determines $\Delta\phi_\mathrm{RL}$. Thus, we argue that the difference in behavior between $\Delta\phi$ and $\Delta\phi_\mathrm{RL}$ arises because changes in barrier gate voltage have a large impact on the breadth of the atomic environment contributing to $\Delta\phi_\mathrm{RL}$, whereas these voltage changes produce much smaller fractional changes in the atomic environment important for $\Delta\phi$.

The extracted magnitudes of the interdot couplings $t_{c}$ and $|\Delta_\mathrm{RL}|$ are shown in Fig.~\ref{FIG3}i. 
As expected, these quantities both increase significantly with $V_\text{BC}$ (by a factor of $\sim$2), consistent with the behavior of the energy gaps in Fig.~\ref{FIG3}f. We have argued throughout that $|\Delta_\mathrm{RL}|$ is important, and Fig.~\ref{FIG3}i makes clear that throughout this range in $V_\text{BC}$ the value of the ratio $|\Delta_\mathrm{RL}|/t_{c} \approx 0.25$, supporting the importance of this interdot-intervalley coupling.

Finally, we use simulations to show that the observed magnitude of the interdot-intervalley coupling $\Delta_\mathrm{RL}$ is consistent with our theoretical expectations. 
(See \cite{SM} for details.)
Comparing $|\Delta_\mathrm{RL}|$ directly to $t_{c}$, as done above for the experimental case, is challenging for theory, because calculating it directly would require more knowledge of the alloy disorder than is available at present.
Fortunately, in \cite{SM} we are able to show that the $|\Delta_\mathrm{RL}|/|\Delta_{i}|$ distributions are independent of the alloy disorder strength and depend only on the shape of the $\ket{\mathrm{L}}$ and $\ket{\mathrm{R}}$ wavefunction orbitals, and thus we focus on those ratios.
We consider a two-dimensional system with alloy disorder, in which an asymmetric DQD confinement potential is generated by three top gates.
The system is first tuned to give a desired tunnel coupling, $t_c$.
We then calculate statistical distributions for the ratio $|\Delta_\mathrm{RL}|/|\Delta_{i}|$, where $i = \tR,\tL$.

In Fig.~\ref{FIG3}j, we plot these probability distributions, computed for the cases of $t_{c} = 56$ and $99~\mu\text{eV}$, which correspond to the first and last $V_\text{BC}$ values (see Fig.~\ref{FIG3}i). 
We note that the distribution for $t_{c} = 99~\mu\text{eV}$ is shifted to higher $|\Delta_\mathrm{RL}|/|\Delta_{i}|$ values compared to that for $t_{c} = 56~\mu\text{eV}$, due to the higher wavefunction overlap in this case. 
In Fig.~\ref{FIG3}j, the experimental values of $|\Delta_\mathrm{RL}|/|\Delta_{i}|$ are shown as vertical dashed and dotted lines for the left and right dots, respectively. 
We find that the experimental values are very close to the modes of the distributions, implying that the experimental findings are fully consistent with our theoretical model. 

\section*{Conclusion}
In this article we presented DAXS spectroscopy of a double quantum dot, resolving all four anticrossings between the ground and first excited states in the two dots. We demonstrated how high resolution spectroscopy of these anticrossings reveals both the intravalley and intervalley tunnel couplings.  Both the amplitudes and the phases of these Hamiltonian parameters were extracted, and connections were made between intuitive physical observables---the anticrossing energy gaps---and the important yet difficult to visualize valley phases. The evolution of the valley couplings was shown to depend on the position of the dots, and we argued that this dependence arises from changes in the overlap of the dot wavefunction with the spatially varying atomic distribution of Ge in the quantum well.

The ability to map out valley phases is an important milestone for current and future experiments in Si/SiGe that depend upon having reliable and stable interdot tunnel couplings and $g$-factors \cite{Woodsg-factor}, since these quantities depend strongly on the valley phase, which we have shown to vary locally in the presence of random-alloy disorder.
In some cases, these local fluctuations may be leveraged to enhance qubit gate operations \cite{Soomro2026preprint}; however, the first step to realizing such benefits is to provide local maps. 
An important next step is therefore to apply DAXS techniques to simultaneously map out valley splitting and valley coupling parameters across wider portions of Si/SiGe quantum wells, to provide independent (i.e., uncorrelated) and statistically significant samplings of these parameters. 
The inclusion of valley phases is crucial for these statistical analyses, since otherwise it is extremely challenging to confirm the existence of deterministically enhanced valley splitting \cite{Woods2025preprint}, which remains a key challenge for Si-based quantum processors.
Such valley phase measurements using DAXS will complement valley phase extraction from $g$-factor measurements \cite{volmer2026mapping,Cywinski2026}, which have been demonstrated in conveyor-mode shuttling. By enabling measurement of additional parameters, DAXS enables the extraction of the both valley phase as well as the magnitude and phase of the interdot-intervalley coupling.

\section*{Acknowledgements}

This material is based primarily upon work supported by the U.S. Department of Energy Office of Science National Quantum Information Science Research Centers as part of the Q-NEXT center. We acknowledge support from Intel Corporation under Cooperative Agreement No.~W911NF-22-2-0037 for providing the device studied here, for the code used to triangulate the quantum dot positions, and for useful discussions, including with Joelle Corrigan and Fahd A.~Mohiyaddin.

\section*{Methods}

The experimental setup is based on our previous demonstration of DAXS~\cite{reilydaxs}. The device is mounted at the mixing chamber plate of a dilution refrigerator with a base temperature of 7 mK. DC control of device gate voltages is supplied with SRS SIM928 floating voltage sources at room temperature, routed to the device through cryogenic loom. Square-wave pulses of MHz bandwidth used for DAXS measurements are applied to the plunger gates $\mathrm{P_L}$ and $\mathrm{P_R}$ through coaxial lines, and these are combined with the DC voltage signals through bias tees mounted on the device PCB. For charge sensing, a low frequency sine wave is applied to the same plunger gates via the cryogenic loom. The resulting charge sensor response is measured using a lock-in amplifier coupled to the charge-sensing quantum dot. For DAXS measurements, the relative amplitudes of the sine wave excitation on each plunger gate are such that the lock-in charge sensor response is maximized along the $\delta$ axis.

Determining lever arms that connect gate voltages to quantum dot energies is performed in a two-step process, because the interdot tunnel couplings required for DAXS are different from those most useful for acquiring the needed lever arms. At each SG voltage, charged-sensed, double-dot bias triangle measurements are acquired with low interdot tunnel coupling. Under this condition, bias triangles are sharp and enable extraction of lever arms. The lever arms change slightly---typically less than 10\%---when the interdot coupling is increased to the DAXS regime. To account for this change, we first measure the charge transition to the left dot at the bias triangle tuning, using the lever arm to extract the electron temperature.  The same measurement then is repeated after increasing the interdot tunnel coupling, analyzing the line-shape in reverse to translate from temperature to the lever arm in the DAXS regime. The right dot lever arm is then chosen by ensuring the polarization line is vertical in a charge stability diagram plotted as a function of $\delta$ and $\epsilon$.

The dot positions in Fig.~\ref{FIG2}i are extracted from an electrostatic model of the Intel Tunnel Falls device that considers the capacitance of the plunger, barrier, and screening gates to the dot, which is modeled as a uniform rectangular distribution of charge. By measuring cross-capacitance between each dot and multiple gates we triangulate the dot position on a grid with 2~nm resolution~\cite{Marcks2025,madzik_operating_2025}. We attribute the location of the right dot underneath $\text{B}_\text{C}$ to the high $\text{B}_\text{C}$ voltage required to sufficiently couple the dots to each other for this experiment combined with alloy disorder effects caused by the 1.7\%~Ge concentration in the quantum well.

Ref.~\cite{SM} presents details on the four-level Hamiltonian used throughout this work, including the relationship between the Hamiltonian parameters in the $\pm z$-valley basis and ground/excited-valley basis. We provide details on our method to fit the DAXS data to the Hamiltonian parameters. The statistical properties of the valley coupling parameters are derived from an effective-mass model. Finally, we explain how the error bars are estimated from two sources of uncertainty: statistical variation in the experimental measurements and effects arising from both higher lying energy levels and changes in the confinement potential shape with changing $\delta$. The latter uncertainties are determined by extensive effective-mass calculations that include alloy disorder and electrostatics from a three-gate geometry.

%\bibliography{References}

%apsrev4-2.bst 2019-01-14 (MD) hand-edited version of apsrev4-1.bst
%Control: key (0)
%Control: author (72) initials jnrlst
%Control: editor formatted (1) identically to author
%Control: production of article title (-1) disabled
%Control: page (0) single
%Control: year (1) truncated
%Control: production of eprint (0) enabled
%

\clearpage

\onecolumngrid

\setcounter{section}{0}
\renewcommand{\thesection}{S\arabic{section}}

% For Supplementary Figures (e.g., S1, S2)
\setcounter{figure}{0}
\renewcommand{\thefigure}{S\arabic{figure}}
\renewcommand{\figurename}{Figure} % Optional: Keeps it as "Figure S1"

% For Supplementary Equations (e.g., S1, S2)
\setcounter{equation}{0}
\renewcommand{\theequation}{S\arabic{equation}}

% For Supplementary Tables
\setcounter{table}{0}
\renewcommand{\thetable}{S\arabic{table}}

\section*{Supplementary Materials}

% % This is still a work in progress... 
\section{Four-level model Hamiltonian and relation between $\pm z$-valley basis and ground/excited-valley basis.}
\label{sec:tunnel-couplings}

Throughout the main text, we employ a four-level model for the single-electron states of the DQD that incorporates a single localized orbital in each dot and a valley degree of freedom. In this Supplementary Materials section, we provide the following:  1) We write down the full matrix form of the four-level model in the $\pm z $-valley basis. 2) We then transform the Hamiltonian into the ground/excited-valley basis. This yields the expressions for intra-valley ($\tilde{t}_{gg}, \tilde{t}_{ee}$) and inter-valley ($\tilde{t}_{ge},\tilde{t}_{eg}$) tunnel couplings in the ground/excited-valley basis that are given in Eqs.~(\ref{tgg})-(\ref{teg}) of the main text. 3) We derive closed-form expressions for the Hamiltonian parameters in the original $\pm z$-valley basis in terms of the $\tilde{t}_{\mu \nu}$ tunnel couplings. 4) Finally, we demonstrate the existence of two parity sectors for the tunnel couplings in the ground/excited-valley basis. In addition, we discuss how a gauge freedom within each parity sector allows us to always choose $0 \leq \Delta \phi = \phi_{\tL} - \phi_{\tR} \leq \pi$ without loss of generality.
%a phase ambiguity that limits us to extracting the magnitudes of valley phases differences.

\subsection{Four-level model in $\pm z$-valley basis}
Our four-level model uses $\{\ket{\tL,+z},\ket{\tL, -z},\ket{\tR,+z},\ket{\tR,-z}\}$ as a basis set, where $\ket{\tL,\tau}$ and $\ket{\tR,\tau}$ are localized in the left and right dots, respectively, and $\tau = \pm z$ is the valley degree of freedom arising from the two-fold degenerate valley minima near the $Z$ point of the band structure of Si \cite{Zwanenburg2013}. Note that we do not include a spin degree of freedom, as no magnetic field is present and spin-orbit coupling is weak in Si. In this localized $\pm z$-valley basis, the DQD Hamiltonian is given by 
\begin{equation}
H=
\begin{pmatrix}
\frac{\varepsilon}{2} & \dL^{*} & t_c & \Delta_{\mathrm{RL}}^{*}\\[2pt]
\dL & \frac{\varepsilon}{2} & \Delta_{\mathrm{LR}} & t_c\\[2pt]
t_c & \Delta_{\mathrm{LR}}^{*} & -\frac{\varepsilon}{2} & \dR^{*}\\[2pt]
\dRL & t_c & \dR & -\frac{\varepsilon}{2}
\end{pmatrix},
\label{eq:H-valley-basis}
\end{equation}
where $\varepsilon$ is the detuning, $t_c$ is the valley-conserving tunnel coupling and $\Delta_{ij} = \mel{i,-z}{H}{j,+z}$ with $i,j \in \{\tL,\tR\}$ are intervalley couplings. The intervalley couplings are complex, $\Delta_{ij} = |\Delta_{ij}|e^{i\phi_{ij}}$, where $\phi_{ij}$ are valley phases. For notational convenience, we use $\Delta_{i} = \Delta_{ii}$ and $\phi_{i} = \phi_{ii}$. 
Note that a microscopic theory relating Eq.~(\ref{eq:H-valley-basis}) to an effective-mass model is given in Sec.~\ref{sec:Construction}. In addition, in Sec.~\ref{sec:Construction} we will discuss the construction of the $\ket{\tL}$ and $\ket{\tR}$ orbital states and establish that $\dRL =\Delta_{\mathrm{LR}}$. Therefore, we only use $\dRL$ throughout. Finally, we assume $t_{c} \geq 0$ without loss of generality. 

\subsection{Transformation of four-level model into the ground/excited-valley basis}

In the main text, we transform the four-level model from the $\pm z$-valley basis (illustrated in Fig. \ref{FIG1}b) to the ground/excited-valley basis (illustrated in Fig. \ref{FIG1}c). Here, we present the details of this transformation that result in the tunnel couplings given in Eqs.~(\ref{tgg})-(\ref{teg}) of the main text.

To begin, note that each $2\times2$ diagonal block of Eq. (\ref{eq:H-valley-basis}) describes a single dot,
\begin{equation}
    H_j=\begin{pmatrix}
    \pm\frac{\varepsilon}{2} & \Delta_j^{*} \\
    \Delta_j & \pm\frac{\varepsilon}{2}
    \end{pmatrix},\label{Hj}
\end{equation}
where $j \in \{\tL,\tR\}$. The ground/excited-valley basis is defined as the basis that diagonalizes Eq.~(\ref{Hj}) for each dot. This can be done with valley eigenstates
\begin{align}
\ket{j,e} &= \frac{1}{\sqrt{2}}\!\left(e^{-i\phi_j/2}\ket{j,+z}+e^{i\phi_j/2}\ket{j,-z}\right), \label{je} \\
\ket{j,g} &= -\frac{i}{\sqrt{2}}\!\left(e^{-i\phi_j/2}\ket{j,+z}-e^{i\phi_j/2}\ket{j,-z}\right), \label{jg}
\end{align}
where $g$ and $e$ denote the ground and excited valleys, respectively. These states are split in energy by the valley splitting $E_{v,j} = 2|\Delta_{j}|$ of dot $j$. These new basis states can represented by the local unitary
\begin{equation}
U_j=\frac{1}{\sqrt{2}}
\begin{pmatrix}
e^{-i\phi_j/2} & -ie^{-i\phi_j/2}\\
e^{i\phi_j/2} & ie^{i\phi_j/2}
\end{pmatrix}.
\end{equation}
The total basis transformation is then given by the block-diagonal unitary $U=U_\tL\oplus U_\tR$, which yields the Hamiltonian in the ground/excited-valley basis,
\begin{equation}
H^{\prime} =  U^\dagger H U =
\begin{pmatrix}
    \frac{\epsilon}{2}+|\dL| & 0 & \tee & \tge \\
    0 & \frac{\epsilon}{2} - |\dL| & \teg & \tgg \\
    \tee & \teg & -\frac{\epsilon}{2} + |\dR| & 0 \\
    \tge & \tgg & 0 & -\frac{\epsilon}{2} - |\dR|
\end{pmatrix}.
\label{eq:Hphase}
\end{equation}
Here,
\begin{align}
  \tgg&= t_c\cos\left(\frac{\Delta\phi}{2}\right)
  -\abs{\dRL}\cos\left(\frac{\Delta\phi}{2}-\Delta\phiRL\right), \label{tggSup}\\
  \tee&= t_c\cos\left(\frac{\Delta\phi}{2}\right)
  +\abs{\dRL}\cos\left(\frac{\Delta\phi}{2}-\Delta\phiRL\right), \\
  \tge&= t_c\sin\left(\frac{\Delta\phi}{2}\right)
  -\abs{\dRL}\sin\left(\frac{\Delta\phi}{2}-\Delta\phiRL\right),\\
  -\teg&=t_c\sin\left(\frac{\Delta\phi}{2}\right)
  +\abs{\dRL}\sin\left(\frac{\Delta\phi}{2}-\Delta\phiRL\right), \label{tegSup}
\end{align}
where $\Delta \phi = \phiL - \phiR$ is the valley phase difference, $\Delta \phiRL = \phiRL - \phiR$, the rows and columns are ordered as $\{\ket{\mathrm{L,e}},\ket{\mathrm{L,g}},\ket{\mathrm{R,e}},\ket{\mathrm{R,g}}\}$, and the Hamiltonian is purely real. The Hamiltonian in the ground/excited-valley basis has intravalley $(\tilde{t}_{gg}, \tilde{t}_{ee})$ and intervalley $(\tilde{t}_{eg}, \tilde{t}_{ge})$ tunnel couplings, where the expressions in Eqs.~(\ref{tggSup})-(\ref{tegSup}) match Eqs.~(\ref{tgg})-(\ref{teg}) given in the main text. Note that these tunnel couplings only depends on valley phase \textit{differences}, as a single valley phase is physically meaningless. In other words, there is an unobservable gauge freedom in which we are free to assign a \textit{single} valley phase to be any arbitrary value of our choosing. In addition, note that the Hamiltonian being purely real is a result of the particular chosen phases in Eq.~(\ref{je}) and Eq.~(\ref{jg}). 

\subsection{Hamiltonian parameters in $\pm z$-valley basis in terms of ground/excited-valley basis parameters}
The Eqs.~(\ref{tggSup})-(\ref{tegSup}) for the tunnel couplings $\{\tilde{t}_{gg}, \tilde{t}_{ee}, \tilde{t}_{ge}, \tilde{t}_{eg}\}$ in the ground/excited-valley basis can be uniquely inverted to yield closed-form expressions for $\{t_c,\abs{\dRL},\Delta\phi,\Delta\phiRL\}$, which are parameters from the Hamiltonian given in Eq.~(\ref{eq:H-valley-basis}) and written in the $\pm z$-valley basis. These expressions are found to be
\begin{align}
  t_c &= \frac{1}{2}\sqrt{\left(\tee+\tgg\right)^{2}+\left(\tge-\teg\right)^{2}},\label{eq:tc}\\
  \abs{\dRL} &= \frac{1}{2}\sqrt{\left(\tee-\tgg\right)^{2}+\left(\tge+\teg\right)^{2}},\label{eq:absDRL}\\
  \Delta\phi &= 2\,\arctan\left(\frac{\tge-\teg}{\tee+\tgg}\right),\label{eq:phiL}\\
  \Delta\phiRL &= \frac{\Delta\phi}{2}+\arctan\left(\frac{\tge+\teg}{\tee-\tgg}\right).
  \label{eq:phiRL}
\end{align}
Therefore, Eqs.~(\ref{eq:tc})-(\ref{eq:phiRL}) allow us to map fitted values of $\{\tilde{t}_{gg}, \tilde{t}_{ee}, \tilde{t}_{ge}, \tilde{t}_{eg}\}$ to $\{t_c,\abs{\dRL},\Delta\phi,\Delta\phiRL\}$, which are the main parameters of interest in this work.

\subsection{Gauge invariant parity sectors of tunnel couplings and Hamiltonian parameter ambiguities} \label{subsec:gauge}

Each valley eigenstates $\ket{j,g/e}$ in Eqs.~(\ref{je})-(\ref{jg}) is an eigenvector of the single-dot Hamiltonian in Eq.~(\ref{Hj}) and is therefore defined only up to an overall phase, i.e. $e^{i\phi} \ket{j,g/e}$ is the same physical state for any phase $\phi$. %Different choices of phases will lead to changes in the phases of the tunnel couplings $\tilde{t}_{\mu \nu}$ in Eqs.~(\ref{tggSup})-(\ref{tegSup}). 
If we want to keep the tunnel couplings real, however, it is sufficient to consider $\phi = 0,\pi$ for each of the four valley eigenstates, i.e $\pm\ket{j,g/e}$. Therefore, we consider $16$ different gauge choices $\{\pm_{1},\pm_{2},\pm_{3},\pm_{4}\}$, which refers to the gauge signs for the $\{\ket{\mathrm{L,e}},\ket{\mathrm{L,g}},\ket{\mathrm{R,e}},\ket{\mathrm{R,g}}\}$ basis states.
Different gauge choices will lead to changes in the signs of the tunnel couplings $\tilde{t}_{\mu \nu}$. For example, $\ket{\tL,e}\to - \ket{\tL, e}$ flips the signs of $\tee$ and $\tge$, while leaving $\teg$ and $\tgg$ unchanged. Therefore, the sign of any individual tunnel coupling $\tilde{t}_{\mu \nu}$ is not gauge invariant.

While the gauge choice can impact the sign of any individual tunnel coupling $\tilde{t}_{\mu \nu}$, it can not change of the magnitude of the tunnel couplings, i.e. $|\tilde{t}_{\mu \nu}|$ is gauge invariant. 
In addition, the product
%Importantly, there does exist one quantity $s$ related to the signs of the tunnel couplings that is gauge invariant. Namely,
\begin{equation}
    s = \operatorname{sign} \bigl(\tee\,\teg\,\tge\,\tgg\bigr) \in \{1,-1\}, \label{eq:sign-sector}
\end{equation}
which is the parity of the number of negative tunnel couplings,
is also gauge invariant. This invariance occurs because any change in gauge flips the sign an even ($0,2,4$) number of $\tilde{t}_{\mu \nu}$. 
Therefore, the gauge invariant content of the tunnel couplings is exactly the four magnitudes $\{|\tilde{t}_{gg}|,|\tilde{t}_{ee}|,|\tilde{t}_{ge}|,|\tilde{t}_{eg}|\}$ together with $s = \pm 1$.

\begin{figure}[tb]
  \centering
  \includegraphics[width=0.6\columnwidth]{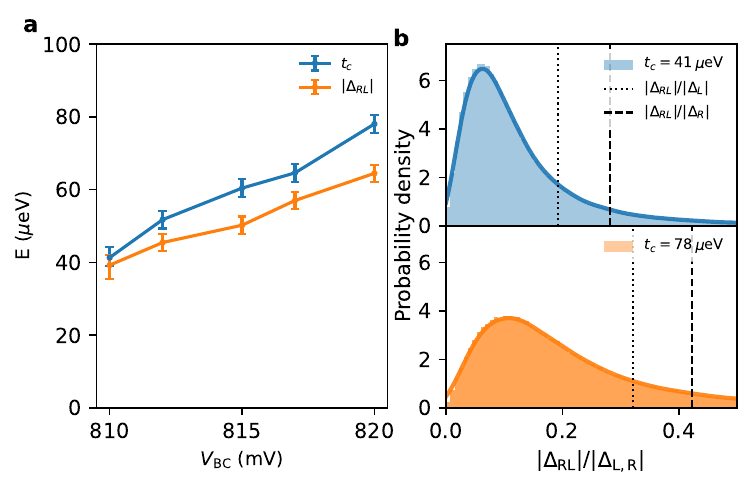}
  \caption{\textbf{Refitting the DAXS results from Fig.~\ref{FIG3} of the main text, except assuming $s = 1$.} {\bf a,} Fitted bare tunnel coupling $t_c$ and interdot-intervalley coupling $\abs{\dRL}$ versus gate voltage $V_{\text{BC}}$. Across the tuning range $\abs{\dRL}$ remains comparable to $t_c$. {\bf b,} Distribution of the ratio $\abs{\dRL}/\abs{\Delta_{\tL,\tR}}$ predicted by the covariance statistics of Sec.~\ref{sec:cov-valley} (shaded), for the lowest and highest $t_c$ tunings in ({\bf a}). Vertical lines mark the extracted ratios $\abs{\dRL}/\abs{\dL}$ (dotted) and $\abs{\dRL}/\abs{\dR}$ (dashed), which lie far in the low-probability tail. The corresponding odd-sector fitting ($s = -1$) is shown in Fig.~\ref{FIG3}i and j of the main text and places these ratios within the bulk of the distribution. Hence, $s = -1$ is more likely, and $s = -1$ is assumed for all results of the main text.}
  \label{FIGS0}
\end{figure}

Within a given $s$ sector, any representative sign assignment of the $\tilde{t}_{\mu \nu}$ may be chosen, and all such choices yield the same spectrum. 
This leads to some ambiguity in mapping an observed spectrum onto the Hamiltonian parameters in the $\pm z$-valley basis.
This is most easily seen by considering how the sign flips of $\tilde{t}_{\mu \nu}$ are carried through Eqs.~\eqref{eq:tc}--\eqref{eq:phiRL}. Suppose that for a given gauge, Eqs.~\eqref{eq:tc}--\eqref{eq:phiRL} yield the parameter set $(t_c, |\dRL|, \Delta \phi, \Delta \phiRL)$. Upon applying other sign assignments that flip an even number of $\tilde{t}_{\mu \nu}$, Eqs.~\eqref{eq:tc}--\eqref{eq:phiRL} will yield four distinct parameter sets,
\begin{align}
  &\text{(i)}\quad \bigl(t_c,\ \abs{\dRL},\ \Delta\phi,\ \Delta\phiRL\bigr),
  \label{eq"gauge-branch1}\\
  &\text{(ii)}\quad \bigl(t_c,\ \abs{\dRL},\ -\Delta\phi,\ -\Delta\phiRL\bigr),
  \\
  &\text{(iii)}\quad \bigl(\abs{\dRL},\ t_c,\ 2\Delta\phiRL-\Delta\phi,\ \Delta\phiRL\bigr),
  \\
  &\text{(iv)}\quad \bigl(\abs{\dRL},\ t_c,\ -(2\Delta\phiRL-\Delta\phi),\ -\Delta\phiRL\bigr).
  \label{eq:gauge-branch4}
\end{align}
Here, (i) comes from the original gauge, flipping the sign of both valley eigenstates of either dot, or flipping the sign of all four valley eigenstates. Case (ii) is obtained by flipping the sign of one valley eigenstate in each dot and coincides with complex conjugation of all valley phases, $\phi_{ij}\to-\phi_{ij}$. Cases (iii) and (iv) are obtained by flipping the sign of either one or three valley eigenstates. Note that $t_{c}$ and $|\dRL|$ are swapped in cases (iii) and (iv). The fact that spectrum of the four-level Hamiltonian is invariant under these parameter changes implies two ambiguities: 1) Our experiment cannot determine the sign of $\Delta \phi$. Therefore, we assume $0 \leq \Delta \phi \leq \pi $ without loss of generality. 2) Our experiment cannot unambiguously determine if $t_{c}$ or $|\dRL|$ is larger. We deal with this ambiguity by assuming that $t_{c} > |\dRL|$. While we can not completely dismiss the possibility that $t_{c} < |\dRL|$, it is much less likely then $t_{c} > |\dRL|$ given the results shown in Fig. \ref{FIG3} of the main text. To see this, suppose that $t_{c}$ and $|\dRL|$ were swapped in Fig. \ref{FIG3}i. We would then have a result in which $|\dRL|/|\dR| \approx 2/3$. This is extremely unlikely given that $|\dRL|/|\dR| = 2/3$ is on the far right of the orange distribution in Fig. \ref{FIG3}j, where $t_{c} = 99~\mu\text{eV}$. This ratio would become even more unlikely for $t_{c} \approx 25~\mu\text{eV}$, which from Fig.~\ref{FIG3}(i) is the appropriate value for $t_{c}$ is the case of $t_{c}$ and $|\dRL|$ being swapped.

Having dealt with the above ambiguities, we still need to decide between $s = \pm 1$. Note that the choice of $s$ is very important, as each possibility of $s$ yields its own values of $\{t_c, \abs{\dRL}, \Delta\phi, \Delta\phiRL\}$ from Eqs.~\eqref{eq:tc}--\eqref{eq:phiRL} and distinct valley splittings $2\abs{\dL}$ and $2\abs{\dR}$ once we fit the data to the four-level model. In principle, the two values of $s$ yield distinct spectra, so the true value of $s$ can be determined by fitting an observed DAXS spectrum to the four-level model. In practice, however, the two values of $s$ yield nearly equally good fits to the data. This is due to the limited range of detuning $\varepsilon$ used in the fitting process. Therefore, fitting to the four-level model is insufficient for inferring $s$. 

However, similar to the choice of gauge above, we can show that $s = -1$ is much more likely than $s = 1$. To see this, we reperform the inference of $|\dRL|$ and $t_{c}$ shown in Fig.~\ref{FIG3}i in the main text, except we assume $s = 1$. The results are shown in Fig.~\ref{FIGS0}a. In contrast to the $s = -1$ results in Fig.~\ref{FIG3}i, $t_c$ and $|\dRL|$ are nearly the same magnitude. Furthermore, $|\dRL|$ for the $s = 1$ results of Fig.~\ref{FIGS0}a is roughly twice as large as the $s = -1$ results in Fig.~\ref{FIG3}i. We then reperform the analysis in Fig.~\ref{FIG3}j for $|\dRL|/|\Delta_{i}|$. The results are shown in Fig.~\ref{FIGS0}b, where the values corresponding to the fitted Hamiltonian parameters are shown by dashed lines. In contrast to the $s = -1$ results in Fig.~\ref{FIG3}i of the main text, we see here that the extracted $|\dRL|/|\Delta_{i}|$ are in the tails of the distributions. Therefore, $s = 1$ is significantly less likely than $s = -1$, and we assume $s = -1$ for all the results of the main text.

\section{Fitting Method} \label{sec:Fitting}

In this Supplementary Materials sections, we provide details on our method to fit the DAXS data to the four-level model described in Figs.~\ref{FIG1}b and c of the main text and Sec.~\ref{sec:tunnel-couplings}. Our fitting method extracts the four tunnel coupling $\{\tgg,\tge,\teg,\tee\}$ in the ground/excited-valley basis and the  valley splittings $\{E_{\text{v,L}}, E_{\text{v,R}}\} = \{2|\Delta_{L}|,2|\Delta_{R}|\}$ of each dot. From the fitted tunnel couplings $\tilde{t}_{\mu \nu}$, we then yield the Hamiltonian parameters $\{t_{c}, |\dRL|, \Delta \phi, \Delta \phi_{\tR\tL}\}$ in the $\pm z$-valley basis (using Eqs.~(\ref{eq:tc})-(\ref{eq:phiRL})) and the anticrossing gaps $E_{\mu \nu}$. 

This section is organized as follows: In Sec.~\ref{sec:UncertaintyModel} we describe how the high-resolution DAXS scans near the anticrossings provide energy-level values and locations, along with establishing uncertainties for these values. In Sec.~\ref{subsec:fit-gapbased} we describe the first step of the fitting process, which uses the anticrossing gaps directly inferred from the high-resolution DAXS data to provide high-quality initial guesses for the Hamiltonian parameters. In Sec.~\ref{subsec:fit-4level} we describe the fitting of the high-resolution DAXS data to the four-level model.
In Sec.~\ref{subsec:fit-mc} we discuss how Monte-Carlo uncertainty propagation is used to propagate the uncertainties in the energy levels established in Sec.~\ref{sec:UncertaintyModel} to uncertainties in the fitted Hamiltonian parameters.

\subsection{Measurement data and determination of energy-level uncertainty} \label{sec:UncertaintyModel}

\begin{figure}[tb]
  \centering
  \includegraphics[width=0.8\columnwidth]{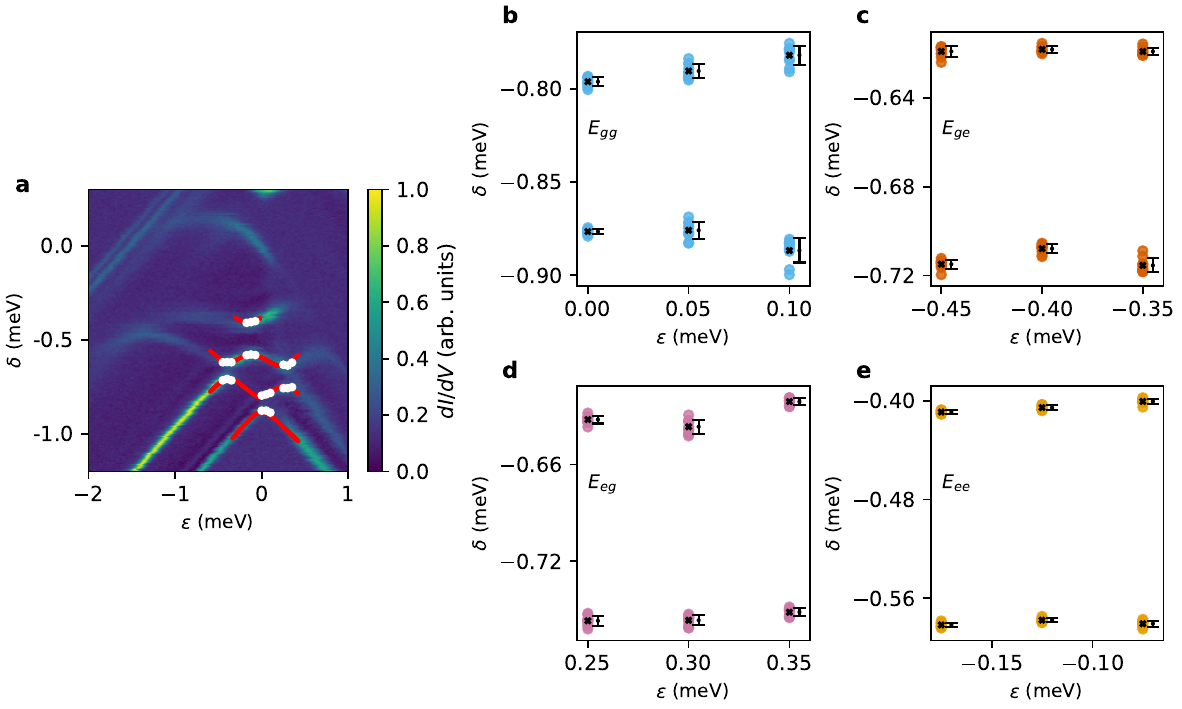}
  \caption{\textbf{Extracting the energy levels of a DAXS spectrum and their uncertainties.} {\bf a,}  A DAXS spectrum yielding a direct measurement of the single-electron energy dispersion, which is the same as Fig.~\ref{FIG2}b of the main text. White markers are the extracted energy levels averaged over $n_{\mathrm{rep}}=10$ repeated high-resolution, one-dimensional DAXS scan along the $\delta$ axis. The red curves are the energies calculated from the four-level Hamiltonian using the parameters from the fitting procedure described in Sec.~\ref{subsec:fit-4level}. {\bf b-e,} Lower and upper energy levels of the four anticrossings, $E_{gg}$, $E_{ge}$, $E_{eg}$, and $E_{ee}$, at the three sampled detunings $\varepsilon$. Colored points are for the individual repeats, the black crosses are the averages (the central values used in the fit), and the error bars are the standard deviations $s_{\mathrm{rep},i}$. The gap $E_{\mu \nu}$ of each anticrossing is the minimum separation between its lower and upper energy level.}
  \label{FIGS1}
\end{figure}

As stated in the main text, for each device tuning, we first acquire a global DAXS spectrum, like the one shown in Fig.~\ref{FIG1}h and reproduced here in Fig.~\ref{FIGS1}a. From this global DAXS spectrum, we determine the approximate locations of the four anticrossings. We then perform high-resolution, one-dimensional DAXS scans for three different $\varepsilon$ values near each anticrossing, as illustrated in Fig.~\ref{FIG2}b of the main text. Each high-resolution scan yields data like shown in Fig.~\ref{FIG2}c of the main text, for which we extract two energy levels by fitting derivatives of the Fermi-Dirac distributions. Note that there are $4 \times 3 \times 2 = 24$ energy levels $(E_i, i = 1,2,\dots ,24)$ for each device tuning, where $4$, $3$, and $2$ come from the number of anticrossings, number of $\varepsilon$ values near each anticrossing, and number of energy levels for each anticrossing, respectively.

Importantly, each high resolution scan is repeated $n_{\mathrm{rep}}=10$ times, which allows us to establish an uncertainty for the extracted energy levels. For each repeated scan, the energy levels are refit. This is illustrated in Fig.~\ref{FIGS1}b-e for the four anticrossings in Fig.~\ref{FIGS1}a, where the colored points are for the $10$ individual energy level fits and the black crosses are the averages. For each energy level, we can calculate a standard deviation $s_{\mathrm{rep},i}$ calculated over the repeated energy-level fits. These are shown by the black error bars in Fig.~\ref{FIGS1}b-e. Each $s_{\mathrm{rep},i}$ then corresponds to a standard error $\sigma_{i}$ of the mean,
\begin{equation}
    \sigma_{i} = \frac{s_{\mathrm{rep},i}}{\sqrt{n_{\mathrm{rep}}}}, \label{eq:peak-sigma}
\end{equation}
for a given energy level. The collection of $\sigma_i$ values of the $24$ energy levels are then used in the Monte-Carlo error propagation in Sec.~\ref{subsec:fit-mc} to determine the uncertainty of the Hamiltonian parameters. Note that the $\sigma_i$ values only capture the \textit{measurement} uncertainty. The reported uncertainties of each extracted parameter in the main text additionally includes a \textit{method} uncertainty quantified by noiseless simulations in Sec.~\ref{sec:val_simul}. These two uncertainties are combined in quadrature as described in Sec.~\ref{subsec:val-method}.

\subsection{Fitting method step 1: gap-based fit}
\label{subsec:fit-gapbased}

In order to generate an estimate for the Hamiltonian parameters, which will serves as an initial guess in the four-level fitting described in Sec.~\ref{subsec:fit-4level}, we use the estimated anticrossing gaps $E^\prime_{\mu \nu}$ and valley splittings $E^\prime_{v,L}$ and $E^\prime_{v,R}$ from the high-resolution DAXS scans described in Sec.~\ref{sec:UncertaintyModel} and illustrated in Fig.~\ref{FIGS1}. Here, we denote $E^\prime_{\mu \nu}$ and $E^\prime_{v,i}$ with a superscript $\prime$ to differentiate them from $E_{\mu \nu}$ and $E_{v,i}$, which are from the four-level model fit described in Sec.~\ref{subsec:fit-4level}.

For each anticrossing, the relevant quantity is the energy gap $E^\prime_{\mu \nu} = \min[E_{\mathrm{high},\mu \nu}(\varepsilon)-E_{\mathrm{low},\mu \nu}(\varepsilon)]$, where $E_{\mathrm{high},\mu \nu}$ and $E_{\mathrm{low},\mu \nu}$ are the high-energy and low-energy levels, respectively, of the anticrossing. Using the three $\varepsilon$ values of the high-resolution DAXS scans for each anticrossing, the squared gap is fit by a parabola, $(E_{\mathrm{high},\mu \nu}(\varepsilon)-E_{\mathrm{low},\mu \nu}(\varepsilon))^2 = a \varepsilon^2 + b \varepsilon + c$, yielding a gap $E^{\prime}_{\mu \nu} = c - b^2/(4a)$ at $\varepsilon_{c} = -b/(2a)$. Note that if $\varepsilon_{c}$ falls outside of the sampled window of $\varepsilon$, we use the smaller gap of the smallest and largest sampled $\varepsilon$. This avoid extrapolation errors. 

For the gap-based fit, we also need the valley splittings $E_{v,L}^\prime$ and $E_{v,R}^\prime$. According to the four-level model, the energies of the centers of the anticrossings are given by
\begin{equation}
    y_{gg} = -\frac{|\dL| + |\dR|}{2}, \quad 
    y_{ge} = \frac{-|\dL| + |\dR|}{2}, \quad
    y_{eg} = \frac{|\dL| - |\dR|}{2}, \quad
    y_{ee} = \frac{|\dL| + |\dR|}{2}. \label{ymunu}
\end{equation}
These equations all hold if the DAXS spectrum perfectly fits the four-level model. In practice, however, the DAXS spectrum will not perfectly satisfy these equations, as there are two variables and three relative energies. If we drop any particular $y_{\mu \nu}$ in Eq.~(\ref{ymunu}), then $|\Delta_{\tL}|$ and $|\dR|$ can be chosen to satisfy the remaining two relative energies. We estimate $|\Delta_{\tL}|$ and $|\dR|$ by considering all four cases, where we drop a different $y_{\mu \nu}$ for each case, and then average the results. Note that obtaining a precise value of $|\dL|$ and $|\dR|$ here is not crucial, as our gap-based fit is only to obtain initial guesses for the four-level model fit described in Sec.~\ref{subsec:fit-4level}.

Finally, we need expressions that relate the anticrossing gaps $E_{\mu \nu}^\prime$ and valley splittings $E_{v,i}^\prime$ to the Hamiltonian parameters. To obtain these, first note that at $\varepsilon = 0$ the diagonal entries of $H^{\prime}$ are the four valley-eigenstate energies $\pm\abs{\Delta_i}$, and the off-diagonal blocks hold the four real tunnel couplings of Eqs.~(\ref{tggSup})--(\ref{tegSup}).
As the detuning $\varepsilon$ is swept, an anticrossing occurs when a left-dot and right-dot valley eigenstate become degenerate. For example $\ket{\mathrm{L,g}}$ and $\ket{\mathrm{R,g}}$ cross at $\varepsilon=\abs{\dL}-\abs{\Delta_\tR}$, with the remaining two states being separated away in energy by the valley splittings $2\abs{\dL}$ and $2\abs{\Delta_\tR}$. 
If the two crossing states were isolated, the Hamiltonian has the form 
\begin{equation}
    H = \begin{pmatrix}
        0 & \tilde{t}_{\mu \nu} \\ 
        \tilde{t}_{\mu \nu} & 0 
    \end{pmatrix}, 
\end{equation} 
which gives $E_{\mu \nu}=2\abs{\tilde{t}_{\mu \nu}}$. Projecting the four-level $H^{\prime}$ onto the two-dimensional subspace with a Schrieffer-Wolff (SW) transformation yields a renormalized anticrossing gap, $E_{\mu \nu} = 2|\tilde{t}_{\mu \nu}^{\text{eff}}|$, where $|\tilde{t}_{\mu \nu}^{\text{eff}}|$ is the effective tunnel coupling given by
\begin{equation}
  \tilde{t}_{\mu \nu}^{\mathrm{eff}} = \tilde{t}_{\mu \nu}^{(1)} + \tilde{t}_{\mu \nu}^{(3)} + \tilde{t}_{\mu \nu}^{(5)} + \cdots,
\end{equation}
where $\tilde{t}_{\mu \nu}^{(n)}$ comes from the $n^\text{th}$-order term of the SW transformation. Note that because the tunnel matrix connects the right dot only to the left dot, every virtual excitation out of the two-dimensional subspace returns to the originating dot only after an even number of hops. The interdot effective coupling therefore receives contributions at odd orders only, and the leading correction to $\tilde{t}_{\mu \nu}$ is $\tilde{t}_{\mu \nu}^{(3)}$. Applying the SW transformation \cite{Winkler2003} to third order then yields
\begin{align}
  \tgg^{\mathrm{eff}}&\approx \tgg+\frac{\tee\teg\tge}{E_{v,\tL}E_{v,\tR}}-\frac{\tgg}{2}\!\left(\frac{\tge^{2}}{E_{v,\tR}^{2}}+\frac{\teg^{2}}{E_{v,\tL}^{2}}\right), \label{eq:sw1}\\
  \teg^{\mathrm{eff}}&\approx \teg-\frac{\tee\tge\tgg}{E_{v,\tL}E_{v,\tR}}-\frac{\teg}{2}\!\left(\frac{\tee^{2}}{E_{v,\tR}^{2}}+\frac{\tgg^{2}}{E_{v,\tL}^{2}}\right),\\
    \tge^{\mathrm{eff}}&\approx \tge-\frac{\tgg\teg\tee}{E_{v,\tL}E_{v,\tR}}-\frac{\tge}{2}\!\left(\frac{\tgg^{2}}{E_{v,\tR}^{2}}+\frac{\tee^{2}}{E_{v,\tL}^{2}}\right),\\
  \tee^{\mathrm{eff}}&\approx \tee+\frac{\tgg\teg\tge}{E_{v,\tL}E_{v,\tR}}-\frac{\tee}{2}\!\left(\frac{\teg^{2}}{E_{v,\tR}^{2}}+\frac{\tge^{2}}{E_{v,\tL}^{2}}\right).
  \label{eq:sw3}
\end{align}
Each correction is suppressed by the ratio of tunnel couplings to valley splitting, $\abs{\tilde{t}_{\mu \nu}}/E_{v,i}$. Therefore, $E_{\mu \nu}\approx 2\abs{\tilde{t}_{\mu \nu}}$ is an accurate estimate of every gap whenever the valley splittings dominate over the tunnel couplings. The third-order terms in Eqs.~(\ref{eq:sw1})-(\ref{eq:sw3}) are the leading refinement when $\tilde{t}_{\mu \nu}$ is extracted from the measured spectrum.
Given the fixed valley splittings $E^{\prime}_{v,i}$ as energy denominators, the four measured gaps $E^\prime_{\mu \nu}$ are inverted through the effective couplings of Eqs.~(\ref{eq:sw1})-(\ref{eq:sw3}), yielding $\tilde{t}_{\mu \nu}$ constrained to odd parity sector $s=-1$, as described in Sec.~\ref{subsec:gauge}. The couplings $\tilde{t}_{\mu \nu}$ are then mapped to $\{t_c,\abs{\dRL},\Delta\phi,\Delta\phiRL\}$ via closed-form inversion using Eqs.~(\ref{eq:tc})--(\ref{eq:phiRL}).

\subsection{Fitting method step 2: four-level model Hamiltonian fit} \label{subsec:fit-4level}
 The final extraction of the Hamiltonian parameters comes from fitting the $24$ peak positions $E_{i}$ ($i = 1,2,\dots,24$) from the DAXS high-resolution scans, as described in Sec.~\ref{sec:UncertaintyModel}, to the spectrum of the four-level Hamiltonian $H^\prime$ given in Eq.~(\ref{eq:Hphase}) with a few small alterations. Let $\lambda_{n}(\varepsilon)$ be the $n^\text{th}$ eigenvalue of $H^\prime$ for detuning $\varepsilon$. The energy values to be fit against are then
 \begin{equation}
     E_{n}(\varepsilon) = \lambda_{n}(\varepsilon - \varepsilon_{0}) + \kappa \left(\varepsilon - \varepsilon_0\right) + E_{\text{off}}.
 \end{equation}
Here, $\varepsilon_{0}$ and $E_{\text{off}}$ are simply detuning and energy offsets that allow for a global translation of the spectrum in the $(\varepsilon,\delta)$-plane. In addition, we include a parameter $\kappa$, which accounts for a tilting of the $\varepsilon$ axis, possibly arising from a miscalibration of the quantum dot lever arms. The total set of parameters is then 
\begin{equation}
  \theta = \left(
  \varepsilon_0,\,E_{\mathrm{off}},\,\kappa,\,
  \abs{\dL},\,\abs{\dR},\,
  \tgg,\,\tge,\,\teg,\,\tee\right).
  \label{eq:fit-params}
\end{equation}
Note that unlike the gap-based fit described in Sec.~\ref{subsec:fit-gapbased}, $\abs{\dL},\abs{\dR}$ are free parameters. The best fit parameters $\theta$ are then determined using a standard least squares method. Note that we use the estimated Hamiltonian parameters from the gap-based fit described in Sec.~\ref{subsec:fit-gapbased} as initial guesses. After determination of $\theta$, the tunnel couplings $\tilde{t}_{\mu \nu}$ are inverted using Eqs.~(\ref{eq:tc})--(\ref{eq:phiRL}) to obtain $\{t_c,\abs{\dRL},\Delta\phi,\Delta\phiRL\}$. Finally, the anticrossing gaps $E_{\mu \nu}$ are determined by calculating the spectrum of $H^\prime$ as a function $\varepsilon$ using the fitted Hamiltonian parameters.

\subsection{Monte-Carlo uncertainty propagation}
\label{subsec:fit-mc}

As described in Sec.~\ref{sec:UncertaintyModel}, the fact that each high-resolution, one-dimensional DAXS scan was repeated $n_{\text{rep}} = 10$ times, allowed us to establish a standard error $\sigma_i$ given in Eq.~(\ref{eq:peak-sigma}) for each energy level $E_{i}$. We propagate these uncertainties to uncertainties in the extracted Hamiltonian parameters using Monte-Carlo uncertainty propagation. 
Specifically, rather than simply using the average energy levels $\bar{E}_i$ (corresponding to the black crosses in Fig. \ref{FIGS1}b-e) in fitting, we fit to the four-level model where each energy level $\bar{E_i}$ is replaced with
\begin{equation}
    E_i = \bar{E}_i + \sigma_i X_i, \label{eq:mc-draw}
\end{equation}
where $\sigma_i$ is the standard error for the energy level $E_i$ and $X_i$ is a random variable drawn from a Student's t distribution with parameter $\nu = n_{\text{rep}} -1 = 9$ \cite{Stdtdist}. For a given DAXS spectrum, we generate $N = 10^4$ such collections of energy levels. We then perform the inference of the Hamiltonian parameters described in Sec.~\ref{subsec:fit-4level} for each collection of energy levels. The range of fitted Hamiltonian parameter can then be specified according to percentile intervals. In Sec.~\ref{subsec:val-method} we describe how we combine these \textit{measurement} uncertainties with \textit{method} uncertainties obtained from device simulations to produce the error bars shown in the main text.

\section{Statistics of intervalley coupling matrix elements from effective mass model} \label{EffMassModel}

Starting from an effective-mass model of a Si/SiGe heterostructure with alloy disorder, we derive in this Supplementary Materials section the statistical correlations between the various intervalley couplings $\Delta_{ij}$. 
The effective-mass model will also be used in Sec.~\ref{sec:val_simul} to estimate the uncertainties in the extracted Hamiltonian parameters arising from projecting out higher-energy excited states beyond the four states used in four-level model.

\subsection{Effective-mass model and effective two-dimensional theory}
We use the effective-mass model described in Refs. \cite{Woodsg-factor,Soomro2026preprint}, except we ignore spin-orbit coupling and do not include a magnetic field. The system Hamiltonian is
\begin{equation}
    H = H_0 + H_v, \label{H0plusHv}
\end{equation}
where $H_0$ and $H_v$ are intravalley and intervalley terms, respectively. The intravalley term is given by
\begin{equation}
     H_0 = \frac{\hbar^2}{2m_t}\left(\hat{k}^2_x+\hat{k}^2_y\right) 
     +\frac{\hbar^2}{2m_l} \hat{k}_z^2
     +V(\BS{r}),
    \label{H0}
\end{equation}
where $\hat{k}_j = -i \partial_{j}$, $m_l = 0.91 m_e$ and $m_{t} = 0.19 m_e$ are the longitudinal and transverse effective masses of Si. Here, the total potential is given by 
\begin{equation}
    V(\BS{r}) = V_{\text{QD}}(\BS{r}) + V_{\text{dis}}(\BS{r}), \label{Potential}
\end{equation}
where $V_{\text{QD}}$ comes from the quantum well confinement potential and applied gate voltages, while $V_{\text{dis}}$ accounts for alloy disorder. The intervalley Hamiltonian is given by
\begin{equation}
    H_{v} = V(\bm{r})e^{+2ik_{0}z}\tau_{+} + V(\bm{r})e^{-2ik_{0}z}\tau_{-},
  \label{eq:Hv}
\end{equation}
where $\tau_{\pm} = \ket{\pm z}\bra{\mp z}$ are the valley raising/lowering operators acting in the $\{\ket{+z},\ket{-z}\}$ pseudospin space, and $2k_0$ is the separation of the two valleys in the Brillouin zone of Si. The rapidly oscillator factors $\exp(\pm i 2k_0z)$ in Eq.~(\ref{eq:Hv}) imply that only the $k_z = \pm 2k_0$ wavevector components of the potential significantly couple the valleys. This corresponds to a wavelength of $2k_0/(2\pi) \approx 0.3~\text{nm}$, which is significantly smaller than typical quantum interface widths~\cite{Wuetz2022p7730}. Therefore, we assume for this work that only $V_{\text{dis}}$ in Eq.~(\ref{Potential}) contributes significantly to the valley coupling \cite{MerrittPracticalVS}.

Following Ref. \cite{Soomro2026preprint}, we approximate $V_{\text{QD}}$ to be separable in the in-plane and growth directions, $V_{\text{QD}}(\BS{r})=  V_{t}(x,y)+V_{l}(z)$. Therefore, the eigenstates of $H_0$ when ignoring alloy disorder are also separable, $\Psi(\BS{r}) = \psi(x,y) \varphi_n(z)$, where $\varphi_{n}$ is the subband wavefunction satisfying
\begin{equation}
     \left[\frac{\hbar^2}{2m_l} \hat{k}_z^2
     +V_l(z)\right] \varphi_{n}(z) = \epsilon_{n} \varphi_{n},
    \label{Hsubband}
\end{equation}
where $\epsilon_n$ is the subband energy. Given that the subband energy splitting $\epsilon_{m} - \epsilon_{n}$ is the dominant energy scale of the system, we can arrive at an effective two-dimensional theory by projecting onto the lowest-energy subband $\varphi_{1}$. This yields the two-dimsional effective Hamiltonian
\begin{equation}
    H_{\text{2D}} = \frac{\hbar^2}{2m_t}\left(\hat{k}^2_x+\hat{k}^2_y\right) 
    + V_{t}(x,y)
    + \Vcal(x,y) 
    + \tilde{\Delta}(x,y)\tau_{-} 
    + \tilde{\Delta}^*(x,y)\tau_{+}, \label{H2D}
\end{equation}
where $\Vcal$ and $\tilde{\Delta}$ are the valley-conserving and intervalley disorder fields, respectively, whose statistics are described in Sec. \ref{sec:disorder-fields} below.

\subsection{Projected disorder fields}
\label{sec:disorder-fields}

The fields $\Vcal(x,y)$ and $\tilde{\Delta}(x,y)$ in Eq.~\eqref{H2D} both descend from a single microscopic three-dimensional alloy-disorder potential $V_{\mathrm{dis}}(\bm{r})$, projected onto the ground longitudinal subband $\varphi_{1}(z)$. In this subsection, we first specify this disorder and calculate its statistical properties.  

To specify the disorder potential, we first discretize the three-dimensional effective-mass Hamiltonian in Eq.~\eqref{H0plusHv} on a tetragonal lattice following Refs.~\cite{Woodsg-factor,Soomro2026preprint,MerrittPracticalVS}, where $a_l$ and $a_t$ are the longitudinal and transverse lattice constants, respectively. The nonfluctuating change in Ge concentration is taken along the growth direction $\hat{z}$, $\nge=\nge(z)$, so that each site is independently Ge with probability $\nge(z)$ and Si with probability $1-\nge(z)$. After absorbing the mean $\Ege\,\nge(z)$ into the longitudinal confinement $V_{l}(z)$, the discretized site-resolved disorder potential is
\begin{equation}
  V_{\mathrm{dis}}(\bm{r}_i) =
  \begin{cases}
    -\Ege\,\nge(z_i), & \text{probability } 1-\nge(z_i)\ \ (\text{Si site}),\\[4pt]
    \Ege\bigl(1-\nge(z_i)\bigr), & \text{probability } \nge(z_i)\ \ (\text{Ge site}),
  \end{cases}
  \label{eq:Vdis}
\end{equation}
where $\Ege=0.6~\mathrm{eV}$ is used to reproduce the conduction band offset between Si and $\text{Si}_{1-x}\text{Ge}_{x}$. By construction it has zero mean and is uncorrelated from site to site,
\begin{align}
  \avg{V_{\mathrm{dis}}(\bm{r})} &= 0, \label{eq:Vmean}\\
  \avg{V_{\mathrm{dis}}(\bm{r})\,V_{\mathrm{dis}}(\bm{r}')}
    &= \delta_{\bm{r},\bm{r}'}\,\Ege^{2}\,\nge(z)\bigl(1-\nge(z)\bigr). \label{eq:Vcorr}
\end{align}

To obtain the statistics of the $\mathcal{V}$ and $\tilde{\Delta}$ fields in Eq.~(\ref{H2D}), we project the disorder potential onto the ground subband $\varphi_{1}$. This yields
\begin{align}
  \Vcal(x_{i},y_{i})
    &= \sum_{j} V_{\mathrm{dis}}(x_{i},y_{i},z_{j})\,\abs{\bar{\varphi}_{1}(z_{j})}^{2},
    \label{eq:V00}\\
  \tilde{\Delta}(x_{i},y_{i})
    &= \sum_{j} V_{\mathrm{dis}}(x_{i},y_{i},z_{j})\,e^{-2ik_{0}z_{j}}\,\abs{\bar{\varphi}_{1}(z_{j})}^{2}
     = \Rcal(x_{i},y_{i}) + i\,\Ical(x_{i},y_{i}),
    \label{eq:D00}
\end{align}
where $i$ and $j$ index the transverse and longitudinal sites and the intervalley field splits into real $\Rcal$ and imaginary parts $\Ical$. The discretized ground-subband  envelopes $\bar{\varphi}_{1}$ are related to their continuum counterparts by
\begin{align}
    % \bar{\psi}(x_{i},y_{i}) &\approx a_{t}\,\psi(x,y), \nonumber\\
    \bar{\varphi}_{1}(z_{j}) &\approx \sqrt{a_{l}}\,\varphi_{1}(z),
\end{align}
which becomes exact as $a_{l}\to0$. 
% In addition, continuum counterpart of discretized in-plane envelope function $\bar{\psi}(x_{i},y_{i}) \approx a_{t}\,\psi(x,y)$, exact as $a_t \to 0$ as longitudinal one. 
Because $V_{\mathrm{dis}}$ has zero mean [Eq.~\eqref{eq:Vmean}], both projected fields inherit it, $\avg{\Vcal}=\avg{\tilde{\Delta}}=0$.
The phase factor $e^{-2ik_{0}z}$ in Eq.~\eqref{eq:D00} produces $[1\pm\cos(4k_{0}z)]$ and $\sin(4k_{0}z)$ terms; these oscillate rapidly on the scale of the envelopes and average towards zero. Dropping them yields the simplified result that the real and imaginary parts have equal variance and are mutually uncorrelated,
\begin{align}
  \avg{\Vcal(x_{i},y_{i}) \, \Vcal(x_{j},y_{j})} &= \delta_{i,j}\,\Ege^{2}\,a_{l}\,\Sigma^{(l)}, \\
  \avg{\Rcal(x_{i},y_{i})\Rcal(x_{j},y_{j})} &= \avg{\Ical(x_{i},y_{i})\Ical(x_{j},y_{j})} = \delta_{i,j}\,\frac{1}{2}\,\Ege^{2}\,a_{l}\,\Sigma^{(l)},
  \label{eq:V00-stats}
\end{align}
where the longitudinal factor is
\begin{equation}
  \Sigma^{(l)}= \int \dd z\,\abs{\varphi_{1}(z)}^{4}\,\nge(z)\bigl(1-\nge(z)\bigr).
  \label{eq:Sigma-l}
\end{equation}
We evaluate $\Sigma^{(l)}$ using a quantum-well Ge profile modeled by a smooth sigmoid,

\begin{equation}
  \nge(z) = \bar{n}_{\mathrm{Ge, Well}} + (\bar{n}_{\mathrm{Ge}} - \bar{n}_{\mathrm{Ge, Well}})\!\left[\,
   1 - \frac{1}{1+\exp\!\bigl(\tfrac{4(-z-d/2)}{w}\bigr)}
  +1 - \frac{1}{1+\exp\!\bigl(\tfrac{4(z-d/2)}{w}\bigr)}\right],
  \label{eq:sigmoid}
\end{equation}
% MY - I have used same values in EDSR paper 
which yields a minimum concentration of $1.7\%$ Ge in the well ($\bar{n}_{\mathrm{Ge, Well}}=0.017$) and a barrier concentration of $30\%$ ($\bar{n}_{\mathrm{Ge}}=0.3$), with interface width $w=1.9~\mathrm{nm}$, well width $d=10.86~\mathrm{nm}$, vertical field $F_{z}=5~\mathrm{mV/nm}$, and band offset $\Ege=0.6~\mathrm{eV}$. For these parameters we obtain 
\begin{equation}
  \Ege^{2}\,a_{l}\,\Sigma^{(l)} \approx 181~\mathrm{meV}^{2},
\end{equation}
which corresponds to a intradot-intervalley coupling standard deviation of $\sigma_{\Delta}\approx 103~\mu\mathrm{eV}$ at an orbital splitting of $1~\mathrm{meV}$.

The fields $\Vcal$ and $\tilde{\Delta}$ characterized here enter the remainder of this work in two distinct ways. First, they are used directly as disorder potentials in the microscopic device simulation of Sec.~\ref{sec:val_simul}, where they are added to the confinement in the two-dimensional Hamiltonian of Eq.~\eqref{H2D}. Second, their matrix elements between localized dot orbitals define the valley couplings of the four-level model. We  construct those orbitals in Sec.~\ref{sec:Construction} and derive the statistics of the resulting couplings in Sec.~\ref{sec:cov-valley}.

The variances above describe the fluctuation at a single atomic site, with per-site variance $\sigma^{2}\equiv\operatorname{Var}(\mathcal{V}_{i})$. When the potential is coarse-grained over a grid cell of area $\Delta x\,\Delta y$ containing $N$ independent atomic sites, the relevant quantity is the cell average $\bar{\Vcal}=\tfrac{1}{N}\sum_{i=1}^{N}\Vcal_{i}$, whose variance is suppressed,
\begin{equation}
  \operatorname{Var}(\bar{\mathcal{V}}) = \frac{1}{N^{2}}\sum_{i=1}^{N}\operatorname{Var}(\mathcal{V}_{i}) = \frac{\sigma^{2}}{N}.
  \label{eq:coarsegrain}
\end{equation}
Eq.~(\ref{eq:coarsegrain}) allows to use a larger lattice spacing in our microscopic device simulations of Sec.~\ref{sec:val_simul} by scaling $\sigma$ as
\begin{equation}
    \sigma^\prime = \frac{a_{t}}{a_t^\prime} \sigma,
\end{equation}
where $\sigma^\prime$ and $a_t^\prime$ refer to the coarse-grained lattice parameters, and $\sigma$ and $a_t$ are for the original atomic-scale lattice.

\subsection{Statistics of intervalley couplings}
\label{sec:cov-valley}

Whereas $\Vcal(x,y)$ and $\tilde{\Delta}(x,y)$ are random \emph{fields} entering the two-dimensional Hamiltonian in Eq.~\eqref{H2D}, the quantities appearing in the four-level model are their matrix elements between the localized orbitals constructed in Sec.~\ref{sec:Construction}: the couplings $\Delta_{j}$ and $\dRL$ of Eqs.~\eqref{DeltajSup}--\eqref{DeltaRLSup} are complex random variables, and it is their joint statistics that we derive here.

The valley coupling of an actual orbital is calculated by weighting the intervalley field $\tilde{\Delta}$ with the in-plane probability density of that orbital. Here, we take the orbital envelopes to be real, without loss of generality. For two arbitrary in-plane envelopes $\psi_a(x,y)$ and $\psi_b(x,y)$, with discretized counterparts $\bar{\psi}_a(x_i, y_i)\approx a_{t}\,\psi_a(x_i,y_i)$, the valley coupling is
\begin{equation}
  \Delta_{ab} = \sum_{i} \bar{\psi}_a(x_{i},y_{i})\bar{\psi}_b(x_{i},y_{i})\,
  \tilde{\Delta}(x_{i},y_{i})
  = \sum_{i} \bar{\psi}_a(x_{i},y_{i})\bar{\psi}_b(x_{i},y_{i})\bigl(\Rcal(x_i, y_i) + i\,\Ical(x_i, y_i)\bigr).
  \label{eq:coupling-gen}
\end{equation}
If $a=b$, it corresponds to a complex number whose magnitude sets the valley splitting $E_{v,a}=2\abs{\Delta_a}$. Since $\tilde{\Delta}$ has zero mean, so does $\Delta_{ab}$. Its real and imaginary parts inherit the site-diagonal statistics of Eq.~\eqref{eq:V00-stats}, i.e. they have equal variance and are mutually uncorrelated. 

Each matrix element is a sum over many independently fluctuating atomic sites, so by the central limit theorem its real and imaginary parts are, to a good approximation, Gaussian. Combined with the equal-variance and vanishing cross-correlation, this makes $\Delta_{ab}$ a circular Gaussian random variable in the complex plane that is centered at the origin. Its magnitude $\abs{\Delta_{ab}}$ therefore follows a Rayleigh distribution~\cite{MerrittPracticalVS}, 
\begin{equation}
    P(\abs{\Delta_{ab}}) = \frac{\abs{\Delta_{ab}}}{\sigma_{ab}^{2}} \exp\left( - \frac{\abs{\Delta_{ab}}^2}{2\sigma_{ab}^2}\right). \label{eq:rayleigh}
\end{equation}
The scale $\sigma_{ab}$ appearing in Eq.~\eqref{eq:rayleigh}, and more generally the full covariance structure of the couplings entering the four-level Hamiltonian, follow from the second moment of Eq.~\eqref{eq:coupling-gen}.
Writing $\Delta_{ab} = \Rcal_{ab} + i\,\Ical_{ab}$ and  combining the envelope weighting of Eq.~\eqref{eq:coupling-gen} with the site-diagonal field statistics of Eq.~\eqref{eq:V00-stats}, the second moment factorizes into longitudinal and transverse factors. Using the continuum
correspondence $\bar{\varphi}_{1}\!\approx\!\sqrt{a_{l}}\,\varphi_{1}$, $\bar{\psi}\!\approx\!a_{t}\,\psi$ (exact as $a_{l},a_{t}\to0$), the equal-variance, mutually uncorrelated structure quoted above takes the explicit form
\begin{align}
  \avg{\Rcal_{ab}\Rcal_{ab}}&= \avg{\Ical_{ab}\Ical_{ab}} = \frac{1}{2}\bigl(\Ege\,a_{t}\sqrt{a_{l}}\bigr)^{2}\,  \Sigma^{(t)}_{ab}\,\Sigma^{(l)}, \\
  \avg{\Rcal_{ab}\Ical_{ab}} &= 0,
  \label{eq:variance}
\end{align}
so that the per-component variance $\sigma_{ab}^{2}\equiv\avg{\Rcal_{ab}\Rcal_{ab}}=\tfrac{1}{2}\bigl(\Ege\,a_{t}\sqrt{a_{l}}\bigr)^{2}\Sigma^{(t)}_{ab}\Sigma^{(l)}$ is precisely the Rayleigh scale in Eq.~\eqref{eq:rayleigh}.
The longitudinal factor is common to all couplings, while the transverse \emph{variance} tensors are set by the in-plane envelopes,
\begin{equation}
  \Sigma^{(t)}_{ab}  = \int \dd x\,\dd y\, (\psi_{a}\psi_{b}\bigr)^{2}.
  \label{eq:Sigma-t-gen}
\end{equation}
The same construction gives the covariance between two distinct valley couplings, 
\begin{align}
  \avg{\Rcal_{ab}\Rcal_{a'b'}}&= \avg{\Ical_{ab}\Ical_{a'b'}}= \frac{1}{2}\bigl(\Ege\,a_{t}\sqrt{a_{l}}\bigr)^{2}\,
  \Theta^{(t)}_{ab,a'b'}\,\Sigma^{(l)}, \\
  \avg{\Rcal_{ab}\Ical_{a'b'}} &= 0,
  \label{eq:covariance}
\end{align}
where the longitudinal factor $\Sigma^{(l)}$ is unchanged. 

For localized orbitals $\psi_\tL,\psi_\tR$, transverse tensors are 
\begin{equation}
  \Sigma^{(t)}_{\tL}   = \int \dd x\,\dd y\, \abs{\psi_{\tL}}^{4},\qquad
  \Sigma^{(t)}_{\tR}   = \int \dd x\,\dd y\, \abs{\psi_{\tR}}^{4},\qquad
  \Sigma^{(t)}_{\tR\tL}  = \int \dd x\,\dd y\, \left(\psi_{\tR}\psi_{\tL}\right)^{2}.
  \label{eq:Sigma-t}
\end{equation}
and the transverse cross tensors are
\begin{align}
  \Theta^{(t)}_{\tL,\tR}   &= \int \dd x\,\dd y\, \abs{\psi_{\tL}}^{2}\abs{\psi_{\tR}}^{2},\label{eq:Theta-LR}\\
  \Theta^{(t)}_{\tL,\tR\tL}  &= \int \dd x\,\dd y\, \abs{\psi_{\tL}}^{2} \bigl(\psi_{\tR}\psi_{\tL}\bigr),\label{eq:Theta-LLR}\\
  \Theta^{(t)}_{\tR,\tR\tL}  &= \int \dd x\,\dd y\, \abs{\psi_{\tR}}^{2} \bigl(\psi_{\tR}\psi_{\tL}\bigr). \label{eq:Theta-RLR}
\end{align}
The full covariance structure of $\{\dL,\dR,\Delta_{\tL\tR},\dRL\}$ is fixed by a single material/longitudinal factor $\Sigma^{(l)}$ and by overlap integrals of the in-plane envelopes $\psi_{\tL},\psi_{\tR}$. Because the tensors are built from products of envelopes that spread over many atomic sites, their magnitude decreases as the wave functions delocalize---a direct consequence of averaging the microscopic disorder over a larger area. The variances $\Sigma^{(t)}_{ab}$ are diagonal special cases of the covariance tensors $\Theta^{(t)}_{ab,a'b'}$, and the real/imaginary decoupling in Eqs.~\eqref{eq:variance} and~\eqref{eq:covariance} implies that the magnitude and phase of each coupling are uncorrelated.

Finally, note that the covariances of all the valley coupling matrix elements have the same prefactor $\frac{1}{2}\bigl(\Ege\,a_{t}\sqrt{a_{l}}\bigr)^{2}\Sigma^{(l)}$ in Eqs.~\eqref{eq:variance}--\eqref{eq:covariance}. Hence, if we consider the ratio of any two valley couplings, this prefactor will cancel with itself. Therefore, the distributions of the dimensionless ratios $\abs{\dRL}/\abs{\dL}$ and $\abs{\dRL}/\abs{\dR}$ are independent of the disorder strength (given by the above prefactor) and are only determined by the wavefunctions $\psi_{\tL}$ and $\psi_{\tR}$, as stated in the discussion of Fig.~\ref{FIG3}i and j in the main text.

\section{Determination of Hamiltonian parameter uncertainties from device simulations}
\label{sec:val_simul}

In the main text, we report error bars on our fitted Hamiltonian parameters. The error bars include contributions from two independent sources. The first source is \textit{measurement} uncertainty in the energy levels, as described in Sec.~\ref{sec:UncertaintyModel}. These uncertainties are propagated to the fitted Hamiltonian parameters by Monte Carlo error propagation, as described in Sec.~\ref{subsec:fit-mc}. The second source is \textit{method} uncertainty. In short, the method uncertainties arise because the four-level model Hamiltonian does not capture all of the physics of the experimental DAXS spectrum. Indeed, there are two main differences between the four-level model Hamiltonian and the experimental DAXS spectrum. First, the DAXS spectrum is influenced by more then just four levels. This is clear upon inspection of a DAXS spectrum, like the one shown in Fig.~\ref{FIG1}h of the main text, where higher-energy states are clearly visible. The couplings to higher-energy orbitals go beyond the capabilities of the four-level model and will cause the true low-energy spectrum to deviate from the four-level Hamiltonian spectrum.
Second, DAXS technically does not precisely measure the spectrum of the quantum dot system. Rather, a DAXS measurement indicates the $\delta$ and $\varepsilon$ values for which a quantum dot energy level crosses the Fermi level of the adjacent reservoirs. While this is closely connected to the quantum dot spectrum, it is not an exact mapping because the shape of the quantum dot confinement potential slightly deforms when $\delta$ changes. In other words, we cannot perfectly shift the potential energy landscape of the quantum dots up and down without also slightly altering its shape.

In this Supplementary Materials section, we quantify these \textit{method} uncertainties through device simulations. Specifically, we model a double quantum dot system using the effective-mass model described in Sec.~\ref{EffMassModel}, where realistic alloy disorder is included and the quantum dot potential arises from applied voltages on top gates. We then calculate theoretical DAXS spectra by scanning gate voltages and determining when an energy level crosses the Fermi level. We then apply the same fitting procedure (described in Sec.~\ref{sec:Fitting}) that has been used for the experiment. By comparing the the fitted results with the exact results taken directly from the simulation, we can quantify the uncertainty of the Hamiltonian parameters due to the intrinsic features of the DAXS method. Importantly, the modeling performed here includes both the effects from the higher-energy orbitals not included in the four-level model and the deformation of the confinement potential when $\delta$ is changed. 

This section is organized as follows: 
In Sec.~\ref{subsec:devicemodel} we described the electrostatic model. In Sec.~\ref{sec:Construction} we describe the construction of the left and right dot orbital states, the calculation of the matrix element in the effective four-level model, and the definition of the $\delta$ and $\varepsilon$ axes. In Sec.~\ref{subsec:val-truth} we describe our calculation of the theoretical DAXS spectrum and explain how our theoretical DAXS spectrum mocks the experimental DAXS measurements. Finally, Sec.~\ref{subsec:val-method} presents our fitting results, method uncertainties, and how the method and measurement uncertainties are combined to yield the error bars of the main text.

\subsection{Device and electrostatic model}
\label{subsec:devicemodel}
Our double dot system is defined by three square top gates, as outlined in Fig.~\ref{FIGS3}a, where the two outer plunger gates have side lengths of $50$ and $30~\text{nm}$, respectively, and the inner barrier gate has a side length of $50~\text{nm}$. This asymmetry is meant to emulate the different orbital energies of the two dots, as visible in the experimental DAXS spectra. The quantum dots are taken to be located in a quantum well $z = 40~\text{nm}$ below the gates. We employ the analytic potential expressions for square gates given in Ref.~\cite{AndersonGateSimul}. The total simulation region has a size of $L_x = L_y = 300~\text{nm}$, and a lattice spacing of $a_t^\prime \approx 4~\text{nm}$ was used in solving the discretized version of the effective-mass Hamiltonian given in Sec.~\ref{EffMassModel}. An example of the electrostatic potential for gate voltages of $\{V_{L}, V_{B},V_{R}\} = \{0.04,0.0145,0.089\}~\text{V}$ is shown in Fig.~\ref{FIGS3}a. These gate voltages are tuned to reproduce the experimental orbital splittings and tunnel couplings. Fig.~\ref{FIGS3}b shows the same potential as Fig.~\ref{FIGS3}a, except an example of the intravalley alloy disorder potential $\mathcal{V}$ has been included, as described in Sec.~\ref{EffMassModel}. Finally, the plunger gate voltages $V_{L}$ and $V_{R}$ serve as independent tuning knobs, while the barrier gate voltage $V_{B}$ is held fixed.

\subsection{Construction of $\delta$ and $\varepsilon$ axes and calculating matrix elements of the effective four-level model} \label{sec:Construction}

\begin{figure}[tb]
  \centering
  \includegraphics[width=0.8\columnwidth]{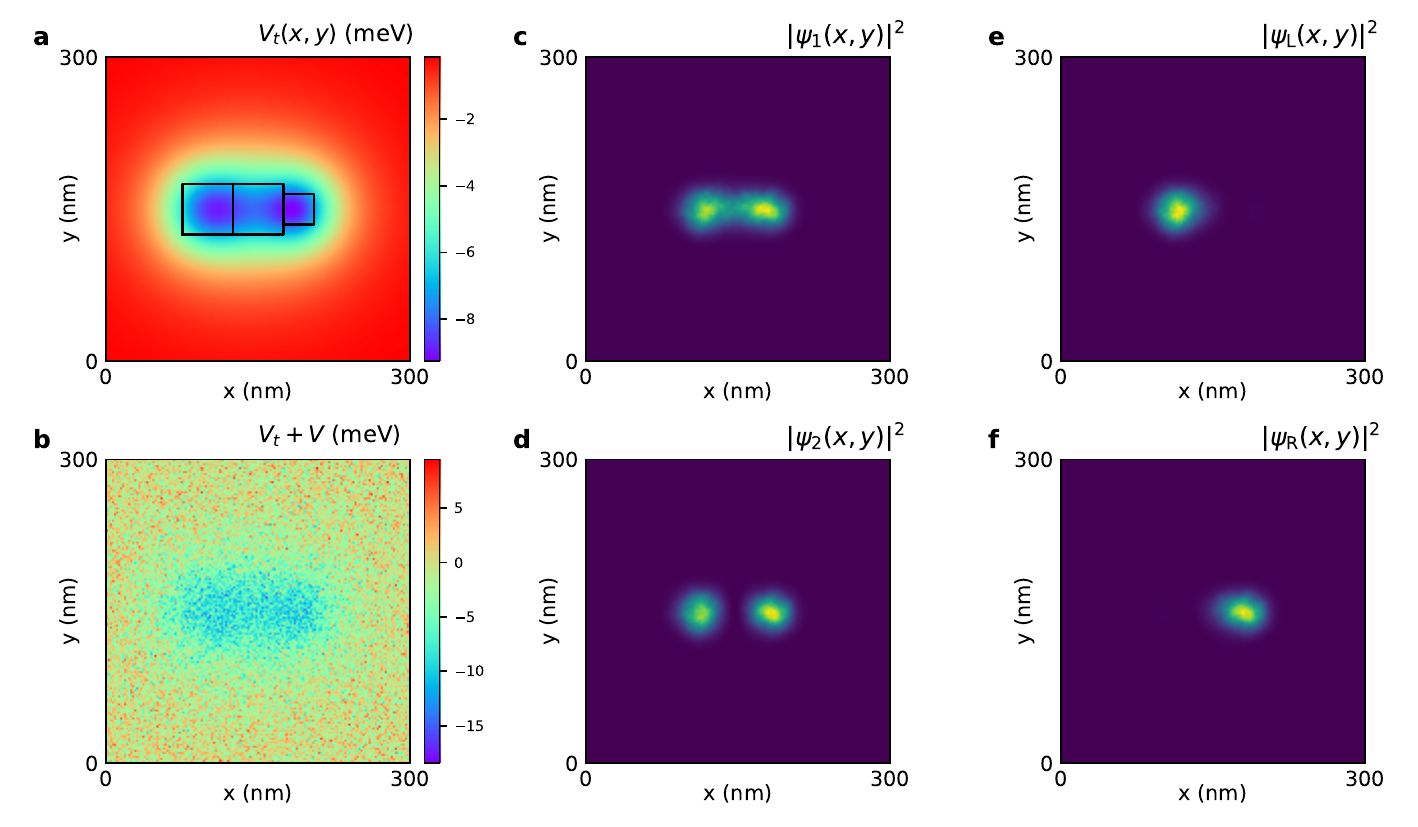}
  \caption{\textbf{Electrostatics, disorder potential, and low-energy dot orbitals.} {\bf a,} In-plane electrostatic confinement $V_{t}(x,y)$ of the three-gate device described in Sec.~\ref{subsec:devicemodel}. Black rectangles outline the gates, with the outer gates being plunger gates and the central gate being a barrier gate. {\bf b,} The same confinement as ({\bf a}) with the coarse-grained intravalley disorder field $\Vcal$ from Eq.~\eqref{eq:V00} added. {\bf c-d,} Probability densities of the two lowest-energy orbital eigenstates of the two-dimensional Hamiltonian in Eq.~\eqref{H2D}. The ({\bf c}) bonding $\abs{\psi_{1}}^{2}$ and ({\bf d}) antibonding $\abs{\psi_{2}}^{2}$ eigenstates are delocalized over both dots. Note that the fuzziness is a result of the disorder. {\bf e-f,} The ({\bf e}) left $\abs{\psi_{\tL}}^{2}$ and ({\bf f}) right $\abs{\psi_{\tR}}^{2}$ localized orbitals constructed from $\psi_{1}$ and $\psi_{2}$ as described in Sec.~\ref{sec:Construction}.}
  \label{FIGS3}
\end{figure}

We now describe how we define the $\delta$ and $\varepsilon$ axes. 
To begin we introduce a parameterization of the transverse potential $V_{t}$ in terms of two plunger gate voltages,  $V_t(x,y) \rightarrow V_{t}(x,y;V_{\text{L}}, V_{\text{R}})$, and decompose $H_{\text{2D}}$ about a reference gate point $(V_{\text{L}}^0,V_{\text{R}}^0)$,  
\begin{equation}
    H_{\text{2D}} = H_{\text{2D,0}}  + H_{\text{2D},v} + H_{\Delta V}, \label{H2D2} 
\end{equation}
where
\begin{equation}
    H_{\text{2D,0}} = \frac{\hbar^2}{2m_t}\left(\hat{k}^2_x+\hat{k}^2_y\right) 
    + V_{t}(x,y;V_{\text{L}}^0,V_{\text{R}}^0)
    + \Vcal(x,y)
\end{equation}
is the intravalley Hamiltonian and includes the valley-conserving disorder potential $\mathcal{V}$, 
\begin{equation}
    H_{\text{2D},v} = \tilde{\Delta}(x,y)\tau_{-} 
    + \tilde{\Delta}^*(x,y)\tau_{+}
\end{equation}
is the intervalley term, and 
\begin{equation}
    H_{\Delta V} = \tilde{V}_{L}h_{\tL} + \tilde{V}_{R}h_{\tR}, \quad h_j = \frac{\partial V_t}{\partial V_j},
\end{equation}
where $\tilde{V}_{j} = {V}_{j} - V_{\text{j}}^0$, and $h_j$ is the lever-arm operator of plunger gate $j\in\{\text{L}, \text{R}\}$. 
We take the reference point $(V_L^0,V_{R}^0)$ to be located on the polarization line, where the electron transitions from being localized in one dot to the other. Note that we do not consider the intervalley coupling $H_{2D,v}$ when defining this polarization line and more details on its definition will be given below. 
Given that we are interested in understanding the physics of the DQD near the reference point, low-energy eigenstates of $H_{\text{2D,0}}$ can be expected to form a good basis. Explicitly, we employ the low-energy states $\psi_{n}$ that satisfy
\begin{equation}
    H_{2D,0} \psi_{n}(x,y)= E_{n,0} \psi_{n}(x,y).
\end{equation}
Furthermore, given that the polarization line corresponds to zero detuning $\varepsilon = 0$, we expect the two lowest-energy states $\psi_{1}$ and $\psi_{2}$ to be delocalized across the DQD and be well separated in energy from the higher-energy excited states, which involve intradot orbital excitations. Therefore, the spatial orbitals $\ket{\psi_1}$ and $\ket{\psi_2}$, when combined with the valley degree of freedom, should form a good basis for small $(\tilde{V}_{L},\tilde{V}_{R})$. We also construct localized orbitals $\ket{L}$ and $\ket{R}$ as
\begin{equation}
    \ket{L/R} = \frac{1}{\sqrt{2}}\left(\ket{\psi_{1}} \pm \ket{\psi_{2}}\right),
\end{equation}
which by construction are degenerate (same energy expectation value) for $\varepsilon = 0$. Examples of $\psi_{1}$ and $\psi_{2}$ and the constructed $\psi_{L}$ and $\psi_{R}$ are shown in Fig.~\ref{FIGS3}c-f.

We can now construct the $\delta$ and detuning $\varepsilon$ axes using the matrix elements of the lever-arm operators. Here, the diagonal elements of $h_{j}$ in the L/R basis are simply the conventional lever arms $\alpha_{ij}$ \cite{leverarm-vanWiel}, which relate changes in the chemical potentials of the two dots to changes in the gate voltages, $\Delta \mu_i = \sum_{j}\alpha_{ij} \tilde{V}_j$, where 
\begin{equation}
    \alpha_{ij} = -\mel{i}{h_{j}}{i}.
\end{equation}
We write $\alpha_{\tL} \equiv \alpha_{\tL\tL}$ and $\alpha_{\tR} \equiv \alpha_{\tR\tR}$ for the response of each dot to its own plunger, and $\alpha_{\tR\tL}$ ($\alpha_{\tL\tR}$) for the cross response of the right (left) dot to the left (right) plunger. Because the dots are inequivalent, $\alpha_{\tL} \neq \alpha_{\tR}$ in general.
The detuning $\varepsilon = \mu_L - \mu_R$ and the polarization line coordinate $\delta = (\mu_\tL + \mu_\tR)/2$ respond to the plungers through lever arms,
\begin{equation}
  \begin{pmatrix} \varepsilon\\ \delta\end{pmatrix}
  =\,\mathcal{A}
  \begin{pmatrix}  \tilde{V}_{\text{L}}\\ \tilde{V}_{\text{R}}\end{pmatrix},
  \qquad
  \mathcal{A}=
  \begin{pmatrix}
    \alpha_\tL-\alpha_{\tR\tL} & \alpha_{\tL\tR}-\alpha_\tR\\
    \frac{1}{2}(\alpha_\tL+\alpha_{\tR\tL}) & \frac{1}{2}(\alpha_{\tL\tR}+\alpha_\tR)
  \end{pmatrix}. \label{eq:eps-del-axes}
\end{equation}
Hence, pure detuning changes require a combination of both plungers, and likewise for pure $\delta$ changes. Inverting Eq.~\eqref{eq:eps-del-axes} gives the virtual gate operators that generate unit motion along each axis, 
\begin{equation}
  h_\varepsilon=\pdv{V_t}{\varepsilon}=\beta^{\varepsilon}_\tL h_\tL+\beta^{\varepsilon}_\tR h_\tR,
  \qquad
  h_\delta=\pdv{V_t}{\delta}=\beta^{\delta}_\tL h_\tL+\beta^{\delta}_\tR h_\tR,
  \label{eq:eps-del-op}
\end{equation}
with coefficient columns are inverse of response matrix, $(\beta^\varepsilon \beta^\delta) = \mathcal{A}^{-1}$.
The detuning operator $h_\varepsilon$ pushes the two dots oppositely, while the $h_\delta$ operator moves them together. Note that its interdot element $\mel{\tR}{h_\delta}{\tL}$ is generically nonzero, which implies that the valley-conserving tunnel coupling drifts along the polarization line, $t_c=t_c(\delta)$. Generically, we may also expect $\mel{\tR}{h_\varepsilon}{\tL}$ to be non-zero. However, we define our reference point $(V_L^0,V_{R}^0)$ to be located on the polarization line only if $\mel{\tR}{h_\varepsilon}{\tL} = 0$. In other words, the valley-conserving tunnel coupling does not change along the $\varepsilon$ axis.
Truncating our basis to the $\ket{L,\pm z}$ and $\ket{R, \pm z}$, then yields the effective four-level Hamiltonian
\begin{equation}
H=
\begin{pmatrix}
\frac{\varepsilon}{2} & \dL^{*} & t_c & \Delta_{\mathrm{RL}}^{*}\\[2pt]
\dL & \frac{\varepsilon}{2} & \Delta_{\mathrm{LR}} & t_c\\[2pt]
t_c & \Delta_{\mathrm{LR}}^{*} & -\frac{\varepsilon}{2} & \dR^{*}\\[2pt]
\dRL & t_c & \dR & -\frac{\varepsilon}{2}
\end{pmatrix} + \tilde{E},
\label{eq:H-valley-basis2}
\end{equation}
whose form matches the four-level model of Eq.~\eqref{eq:H-valley-basis} with parameters given by
\begin{align}
    t_{c} &= \mel{R,\tau}{H_{2D,0}}{L,\tau} + \mel{R}{h_{\delta}}{L} \cdot \delta, \label{tcSup} \\
    \Delta_{j} &= \mel{j,-z}{H_{2D,v}}{j,+z}, \label{DeltajSup} \\
    \dRL &= \mel{R,-z}{H_{2D,v}}{L,+z}, \label{DeltaRLSup}
\end{align}
and $\tilde{E} = (E_{1,0} + E_{2,0})/2$ is average orbital energy of $\psi_{1}$ and $\psi_{2}$ at $\varepsilon = 0$.
The one important difference between Eq.~\eqref{eq:H-valley-basis2} and the four-level Hamiltonian given in Eq.~\eqref{eq:H-valley-basis} and used for fitting to the DAXS spectrum (as described in Sec.~\ref{sec:Fitting}) is that $t_c$ in Eq.~\eqref{tcSup} depends on $\delta$, i.e. the position along the delta axis. This change in $t_c$ with $\delta$ is precisely arising from the shape of the confinement potential changing with $\delta$, as highlighted above. Finally, note that $\Delta_{LR} = \Delta_{RL}$ follows from the fact that $\psi_{1}$ and $\psi_{2}$ can be chosen (without loss of generality) to be real valued.

\begin{figure}[tb]
  \centering
  \includegraphics[width=1\columnwidth]{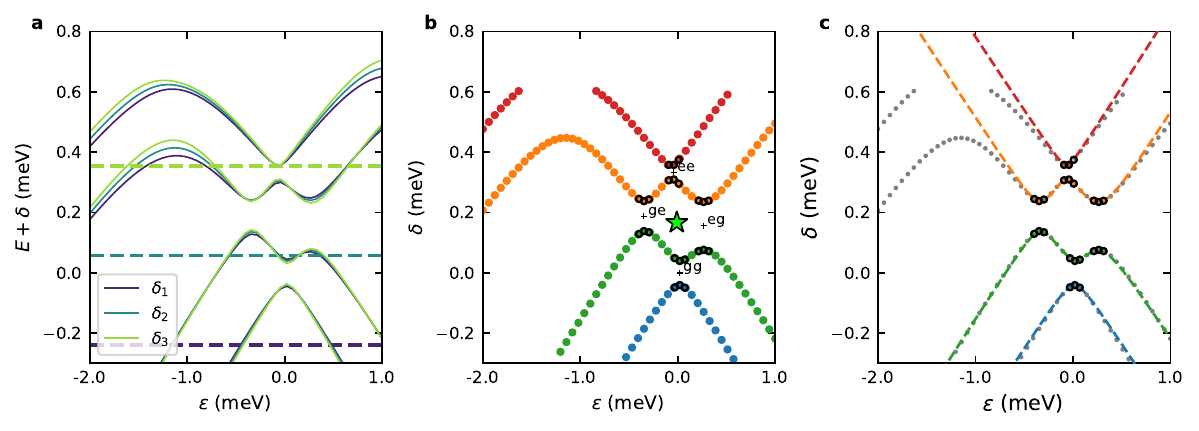}   
  \caption{\textbf{Theoretical DAXS spectrum.}
  {\bf a,} Energy-level spectrum as a function of detuning $\varepsilon$ for three $\delta$, $\delta_1 = -0.24,\;\delta_2 = 0.06,\;\delta_3 = 0.35$ meV (color coded), plotted with ($E+\delta$). Note that for increasing $\delta$, the anticrossing gaps are slightly decreasing due to the deepening of the dot potential wells relative to the barrier between the dots.
  Dashed lines mark the Fermi level $E_{\mathrm{F}}+\delta$ for each $\delta$: as $\delta$ increases, the Fermi energy moves up through the spectrum, so each anticrossing meets the Fermi level at its own $\delta$.
  {\bf b,} Theoretical DAXS spectrum. Colored dots indicates $(\varepsilon,\delta)$ values for which an energy level crosses the Fermi level $E_F$.
  Black circles are the $24$ sampled energy levels ($4$ anticrossings $\times$ $3$ detunings $\times$ $2$ energy levels) that mock the energy level extraction in the experiment (see Sec.~\ref{sec:UncertaintyModel}). Crosses mark the four anticrossing centers, which form a parallelogram. The green star marks its center $(\epsilon_{c},\delta_{c})$, the point at which the matrix elements are evaluate to determine the true Hamiltonian parameters. {\bf c,} Fit for the same DAXS spectrum as ({\bf b}). The DAXS spectrum (gray) taken from ({\bf b}) and the four-level model spectrum (colored dashed) evaluated using $H^{\prime}$ in Eq.~\eqref{eq:Hphase} with the fitted Hamiltonian parameters. The fitted spectrum reproduces the DAXS spectrum within the anticrossing regions and departs from the DAXS spectrum at large $\abs{\varepsilon}$, where the omitted higher orbitals bend the spectrum. This illustrates that the four-level description is valid in the neighborhood of the anticrossings that the fitting procedure samples.}
  \label{FIGS4}
\end{figure}

\subsection{Theoretical DAXS spectrum and mock measurement and parameter fitting} \label{subsec:val-truth}

Having established the polarization line and the $\delta$ and $\varepsilon$ axes for any given point on the polarization line, we are in position to calculate a theoretical DAXS spectrum that mimics the experimental DAXS spectra of this work. This is done as follows: First, for a given disorder realization, we find a point on the polarization line by sweeping $V_{L}$ and $V_{R}$ until we find a point that satisfies the polarization line conditions given in Sec.~\ref{sec:Construction}. Note that the polarization line in the experiment is traced along the ground-ground valley anticrossing. Here, however, it follows the construction of Sec.~\ref{sec:Construction}, where the polarization line is defined without accounting for valleys. The two definitions agree on the directions of the $\delta$ and $\varepsilon$ axes, as the valley states of a dot share the same orbital envelope and hence, identical lever arms. Therefore, the valley field only offsets the crossing positions within the $(\varepsilon,\delta)$-plane without tilting the polarization line. Second, the Fermi level $E_F$ is set using the energy-level spectrum of the reference point. Explicitly, we set $E_F = \tilde{E} - (|\Delta_{L}| + |\Delta_{R}|)/2$, which is the energy in the middle of the ground-ground anticrossing of the effective four-level Hamiltonian in Eq.~\eqref{eq:H-valley-basis2}. 
Third, we then sweep $(\varepsilon,\delta)$ and find the low-energy states of the full two-dimensional Hamiltonian (Eq.~\eqref{H2D2}). We label these energies by $E_n(\varepsilon,\delta)$. An example of how the spectrum changes for three values of $\delta$ is shown in Fig.~\ref{FIGS4}a. Notice that for increasing $\delta$, the anticrossing gaps in Fig.~\ref{FIGS4}a are slightly decreasing. Physically, this makes sense, since the potential minima beneath the plunger gates go down with increasing $\delta$ faster than the local potential maximum underneath the barrier gate, leading to $t_c$ decreasing for increasing $\delta$. Fourth, we determine points in the $(\varepsilon,\delta)$-plane where a state crosses the Fermi level, $E_{n}(\varepsilon,\delta) = E_{F}$. An example of this is shown in Fig.~\ref{FIGS4}b, and we refer to this as a theoretical DAXS spectrum. These are the points that would show up in the signal of an experimental DAXS spectrum. 
We stress that the values in Fig.~\ref{FIGS4}b come from the full microscopic Hamiltonian, and therefore retain complete orbital and valley hybridization and the effects of deforming the shape of the potential with changing $\delta$. 

Having a DAXS spectrum, we now extract the same $24$ energy levels ($4$ anticrossings $\times$ $3$ detunings per anticrossing $\times$ $2$ energy levels per anticrossing) provided by the experiment, as described in Sec.~\ref{sec:UncertaintyModel}. An example of these points are shown in Fig.~\ref{FIGS4}b by black circles. Furthermore, we run the same fitting procedure (as described in Sec.~\ref{sec:Fitting}) that is ran on the experimental data. This yields fitted Hamiltonian parameters $\{t_c, |\Delta_{RL}|, \Delta \phi, \Delta \phi_{RL}\}$, just like the analysis of the experimental data. The spectrum of the four-level Hamiltonian with the fitted Hamiltonian parameters is shown in Fig.~\ref{FIGS4}c by colored dashed lines. We see remarkable agreement with the theoretical DAXS spectrum in the region of the anticrossings, shown in Fig.~\ref{FIGS4}c by the gray dotted lines. For $\varepsilon$ past the anticrossings, the four-level spectrum and theoretical DAXS spectrum deviate due to the hybridization with excited orbitals. This is completely unimportant for fitting the Hamiltonian parameters, however, as the fitted data all comes from near the four anticrossings, where the four-level model well describes the physics.

Importantly, we can compare the fitted Hamiltonian parameters of our simulations to the known Hamiltonian parameters taken directly from evaluating the relevant matrix elements. The only subtlety is that the matrix elements slightly vary over the $(\varepsilon,\delta)$-plane in the simulation. Therefore, we must pick a definite point from which we extract the matrix elements. We choose the point precisely in the center of the parallelogram defined by the four anticrossings, as shown by the green star in Fig.~\ref{FIGS4}b. The idea behind this choice is that $t_{c}$ changes with $\delta$, with $t_{c}$ being slightly smaller at the excited-excited valley anticrossing than the ground-ground valley anticrossing, for example. Hence, a good representative value of $t_c$ should happen at an intermediate $\delta$ value, such as at the green star.

\begin{figure}[tb]
  \centering
  \includegraphics[width=0.8\columnwidth]{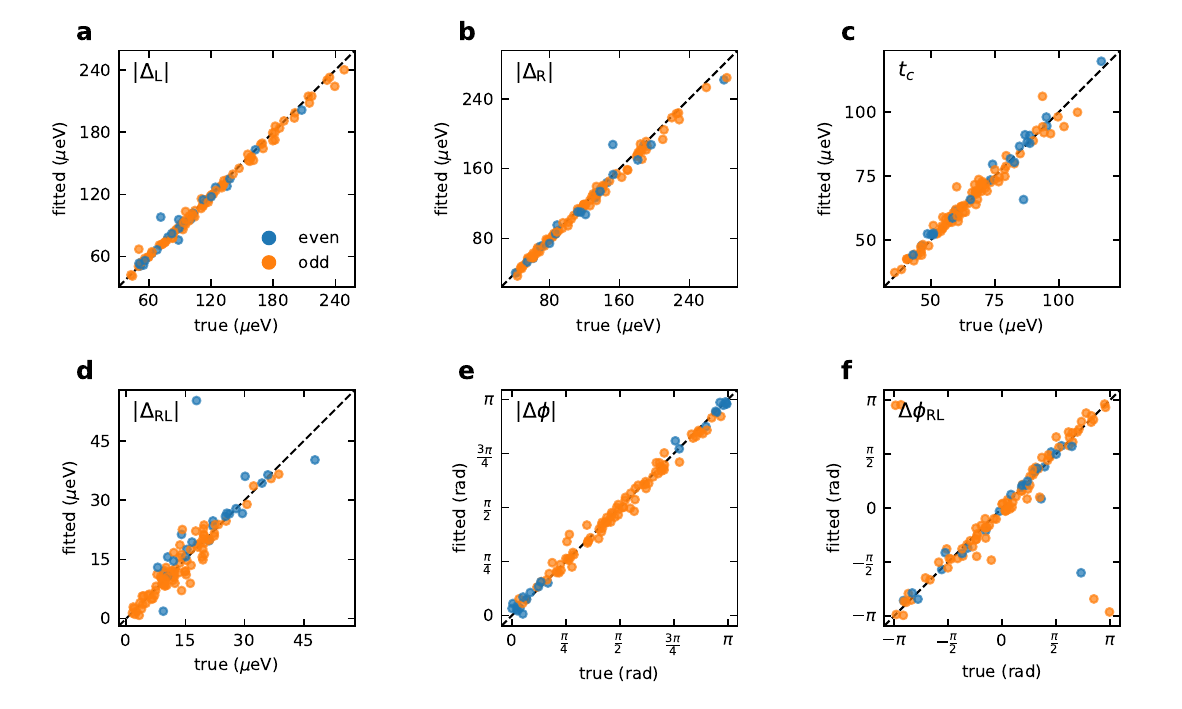}
    \label{fig:s-sim-benchmark}
    \caption{\textbf{Comparison of fitted Hamiltonian parameters and true parameters from simulation.} Fitted versus true values of the six Hamiltonian parameters returned by the four level fitting produce (described in Sec.~\ref{sec:Fitting}). Data is shown for $128$ disorder realizations. {\bf a-b,} left (right) valley coupling strength $\abs{\dL}$ ($\abs{\Delta_\tR}$), {\bf c,} bare tunnel coupling $t_c$, {\bf d,} interdot-intervalley coupling magnitude $\abs{\dRL}$, {\bf e,} valley phase difference $\Delta\phi$, and {\bf f,} interdot-intervalley phase $\Delta\phiRL$. Each point is one disorder realization and the dashed line is the identity $y=x$ (zero error). Points are colored by the tunnel coupling parity $s$ described in Sec.~\ref{subsec:gauge}, with odd ($s=-1$) in orange and even ($s=+1$) in blue. The magnitudes and phases all fall close to identity line over the full range.}
    \label{FIGS5}
\end{figure}

\subsection{Fitting results and method uncertainties}
\label{subsec:val-method}

A comparison of the fitted Hamiltonian parameters to the true parameters for $128$ disorder realizations is shown in Fig.~\ref{FIGS5}. Note that we carried out the procedure for $150$ disorder realizations, but disregarded realizations in which $\abs{\Delta_{i}} <40~\mu\mathrm{eV}$ for at least one dot. In these cases, where the tunnel coupling dominates over the low valley splittings, the individual anticrossings can not be resolved, which is not the case for the experimental results of this paper. All six Hamiltonian parameters cluster tightly along the identity line ($y=x$), which represents zero error. The bare tunnel coupling $t_c$ and the intradot-intervalley couplings $\abs{\dL},\abs{\dR}$ are recovered to a median relative error of $\sim 2\%$, and the valley phases to $\sim 0.1$~rad. The interdot-intervalley coupling $\abs{\dRL}$ shows a larger median relative error ($\sim 14\%$) and correspondingly broader scatter in Fig.~\ref{FIGS5}d. 

\begin{figure}[tb]
  \centering
  \includegraphics[width=0.8\columnwidth]{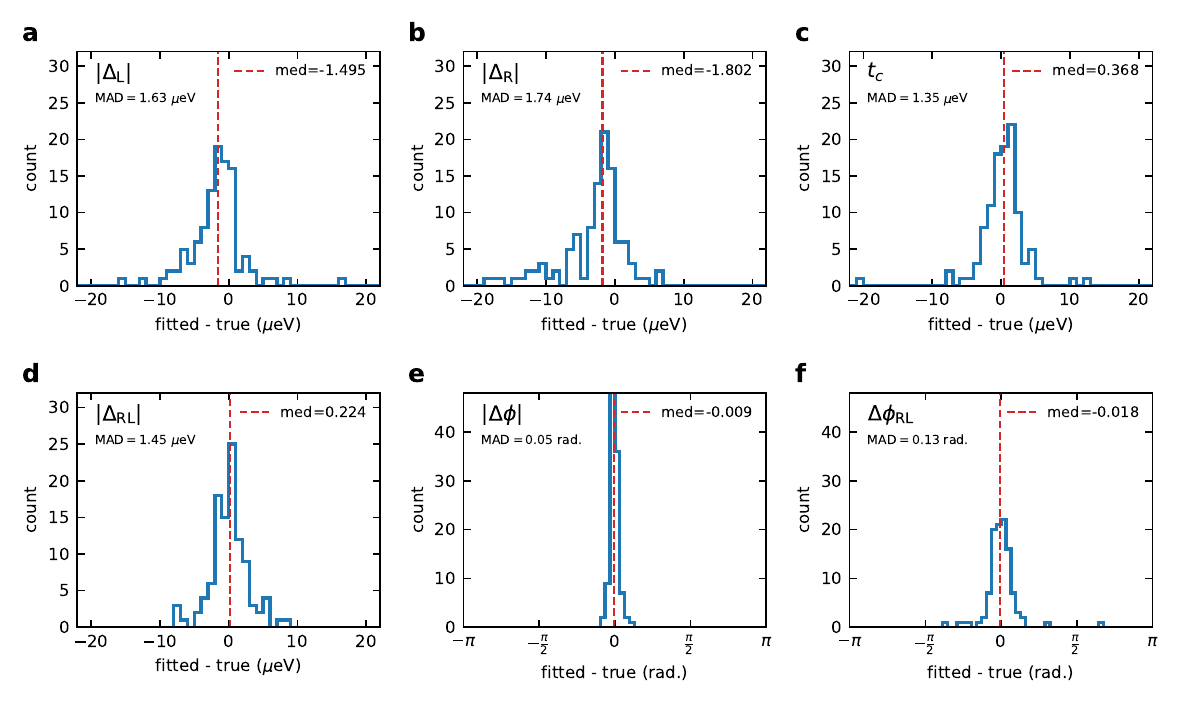}
  \caption{\textbf{Residual distributions of the fitting method against device simulation.} Signed fit residual (fitted $-$ true) for each of the six Hamiltonian parameters, over the $128$ disorder realizations, with panels ({\bf a-f}) as in Fig.~\ref{FIGS5}. In each panel the dashed red line marks the median residual (med) and MAD is the median absolute deviation about it. Every distribution is sharply peaked and roughly symmetric about zero, with a median within $\sim 2~\mu\mathrm{eV}$ (or $\sim 0.015$~rad) of zero, confirming that the fitting is essentially unbiased for all parameters.}
  \label{FIGS6}
\end{figure}

The signed residuals $e=\theta_{\mathrm{fit}}-\theta_{\mathrm{true}}$ shown in Fig.~\ref{FIGS6} make it clear that there are no significant systematic biases in the parameter fitting. Every distribution is sharply peaked, and the tunnel coupling $t_{c}$, the interdot-intervalley coupling $\abs{\dRL}$, and the valley phases are essentially unbiased ($\abs{\operatorname{median}(e)}\lesssim 0.4~\mu\mathrm{eV}$ and $\lesssim 0.02$~rad). The phase residuals have MADs of $0.05$--$0.15$~rad, with $\Delta\phiRL$ broader than $\Delta\phi$, as expected when $\abs{\dRL}$ is small and its phase becomes ill-defined. The only appreciable offsets are in the intradot- intervalley couplings, $\operatorname{median}(e)\approx-1.5~\mu\mathrm{eV}$ for $\abs{\dL}$ and $\approx-1.8~\mu\mathrm{eV}$ for $\abs{\dR}$, the residual signature of the states omitted by the four-level truncation. The residual scatter sets the method's systematic floor, which we quantify next.

The residual distribution of Fig.~\ref{FIGS6} quantify the uncertainties intrinsic to the four-level model fitting described in Sec.~\ref{sec:Fitting}.  
We summarize the spread of each distribution by its median absolute deviation,
\begin{equation}
  \mathrm{MAD} = \operatorname{median}\,\bigl(\,\abs{e_i - \operatorname{median}(e)}\,\bigr).
  \label{eq:mad}
\end{equation}
Assuming the bulk of each distribution is approximately Gaussian, we convert the MAD to an equivalent standard deviation by the known coefficient, $\sigma_{\mathrm{rob}} = 1.4826\,\mathrm{MAD}$. Since the MAD is taken about the median it captures only the random fluctuation and excludes any systematic offset, Thus, to retain that offset without correcting the central value, we take the method uncertainty to be
\begin{equation}
  \sigma_{\mathrm{method}} = 1.4826\,\mathrm{MAD} + \abs{\operatorname{median}(e)},
  \label{eq:sigma-method}
\end{equation}
which is the scale plus the magnitude of the bias. The tunnel coupling $t_{c}$, the interdot valley coupling $\abs{\dRL}$, and the valley phases are essentially unbiased, so this term is negligible for them. The intradot-intervalley couplings $\abs{\dL}$ and $\abs{\dR}$ carry systematic offsets of $\approx-1.5$ and $-1.8~\mu\mathrm{eV}$, respectively, as shown in Fig.~\ref{FIGS6}a and b. We deliberately do not bias-correct the central values---doing so would assume the simulated offset transfers exactly to the experiment---and instead carry $\abs{\operatorname{median}(e)}$ symmetrically in the uncertainty (Eq.~\eqref{eq:sigma-method}). This raises the $\abs{\dL},\abs{\dR}$ method terms by $\sim55$--$65\%$ over the pure scale, which is the conservative choice and keeps the unadjusted fit as the central value while the reported interval still covers the offset.

This method uncertainty is statistically independent of the experimental measurement uncertainty, which we obtain separately from the Monte-Carlo propagation of the peak position standard errors (see Sec.~\ref{subsec:fit-mc}). Because the simulation that determines $\sigma_{\mathrm{method}}$ is noiseless, the two sources share no common variance and are combined in quadrature. The Monte Carlo returns an asymmetric $68\%$ interval $[l,u]$ about the central fit $c$ (the 16th and 84th percentiles of the resampled fits), we fold in the method term independently on each side so as to preserve this asymmetry, giving the reported bounds
\begin{equation}
  l' = c - \sqrt{(c-l)^2 + \sigma_{\mathrm{method}}^2},\qquad
  u' = c + \sqrt{(u-c)^2 + \sigma_{\mathrm{method}}^2}.
  \label{eq:combined-bounds}
\end{equation}
Across all parameters $\sigma_{\mathrm{method}}$ exceeds the measurement term by factors of $\sim 3$--$10$, so the precision of the extracted Hamiltonian parameters is limited by the method uncertainties rather than by measurement uncertainties. The intervals $[l',u']$ of Eq.~\eqref{eq:combined-bounds} are those quoted for the extracted parameters in the main text.

\end{document}